\documentclass[letterpaper, 10 pt, conference]{ieeeconf}

\IEEEoverridecommandlockouts                              
\usepackage{mathtools}    
\usepackage{amsfonts}
\usepackage{graphicx}   
\usepackage{subcaption}
\usepackage{epsfig} 
\usepackage{cancel}
\usepackage{amssymb}
\usepackage{color}
\usepackage{bm}
\usepackage{todonotes} 
\usepackage{float}
\usepackage{cite}
\usepackage{subcaption}
\usepackage{siunitx}
\usepackage[ruled,vlined,titlenotnumbered,linesnumbered]{algorithm2e}

\newcommand{\R}{\mathbb{R}} 
\newcommand{\state}{x} 
\newcommand{\ctrl}{a} 
\newcommand{\dstb}{b} 
\newcommand{\traj}{\zeta}

\newcommand{\fdyn}{f} 
\newcommand{\cset}{\mathcal{A}} 
\newcommand{\cfset}{\mathbb{A}} 
\newcommand{\dset}{\mathcal{B}} 
\newcommand{\dfset}{\mathbb{B}} 

\newcommand{\cost}{J} 
\newcommand{\valfunc}{G} 
\newcommand{\brs}{\mathcal{G}} 
\newcommand{\frs}{\mathcal{W}} 
\newcommand{\targetset}{\mathcal{G}_0} 
\newcommand{\ham}{H} 
\newcommand{\ic}{g} 
\newcommand{\costate}{\lambda}

\title{\LARGE \bf Hamilton-Jacobi Reachability: A Brief Overview and Recent Advances}
\author{Somil Bansal*, Mo Chen*, Sylvia Herbert* and Claire J. Tomlin
\thanks{* All authors contributed equally to this article. Authors' names are written in the alphabetical order. All authors are with the Department of Electrical Engineering and Computer Sciences, University of California, Berkeley. \{somil, mochen72, sylvia.herbert, tomlin\}@eecs.berkeley.edu}
\thanks{This tutorial is supported by NSF under the CPS Frontiers VehiCal project (1545126) and CPS:ActionWebs (CNS-931843), by the UC-Philippine-California Advanced Research Institute under project IIID-2016-005, by the ONR MURI Embedded Humans (N00014-16-1-2206), and by NASA under grants NNX12AR18A and UCSCMCA-14-022 (UARC).}
}

\begin{document}
\maketitle
\thispagestyle{empty}
\pagestyle{empty}

\begin{abstract}
Hamilton-Jacobi (HJ) reachability analysis is an important formal verification method for guaranteeing performance and safety properties of dynamical systems; it has been applied to many small-scale systems in the past decade. Its advantages include compatibility with general nonlinear system dynamics, formal treatment of bounded disturbances, and the availability of well-developed numerical tools. The main challenge is addressing its exponential computational complexity with respect to the number of state variables. In this tutorial, we present an overview of basic HJ reachability theory and provide instructions for using the most recent numerical tools, including an efficient GPU-parallelized implementation of a Level Set Toolbox for computing reachable sets. In addition, we review some of the current work in high-dimensional HJ reachability to show how the dimensionality challenge can be alleviated via various general theoretical and application-specific insights.
\end{abstract}

\section{Introduction \label{sec:introduction}}
As the systems we design grow more complex, determining whether they work according to specification becomes more difficult.
Consequently, verification and validation have received major attention in many fields of engineering.
However, verification of systems is challenging for many reasons. 
First, all possible system behaviors must be accounted for. This makes most simulation-based approaches insufficient, and thus formal verification methods are needed. 
Second, many practical systems are affected by disturbances in the environment, which can be unpredictable, and may even contain adversarial agents. 
In addition, these systems often have high dimensional state spaces and evolve in continuous time with complex, nonlinear dynamics.

Hamilton-Jacobi (HJ) reachability analysis is a verification method for guaranteeing performance and safety properties of systems, overcoming some of the above challenges. 
In reachability analysis, one computes the reach-avoid set, defined as the set of states from which the system can be driven to a target set while satisfying time-varying state constraints at all times. 
A major practical appeal of this approach stems from the availability of modern numerical tools, which can compute various definitions of reachable sets \cite{Sethian96, Osher02, Mitchell02, Mitchell07b}. 
For example, these numerical tools have been successfully used to solve a variety of differential games, path planning problems, and optimal control problems. 
Concrete practical applications include aircraft auto-landing \cite{Bayen07}, automated aerial refueling \cite{Ding08}, model predictive control (MPC) of quadrotors \cite{Bouffard12,Aswani2013}, multiplayer reach-avoid games \cite{Huang11}, large-scale multiple-vehicle path planning \cite{Chen2016d, Chen15b}, and real-time safe motion planning \cite{Herbert2017}.
However, HJ reachability becomes computationally intractable as the state space dimension increases. 
Traditionally, reachable set computations involve solving an HJ partial differential equation (PDE) on a grid representing a discretization of the state space, resulting in an \textit{exponential} scaling of computational complexity with respect to system dimensionality; this is often referred to as the ``curse of dimensionality.''
However, recent work has made a significant leap in overcoming these challenges by exploiting system structures to decompose the computation of reachable set into several small dimensional computations \cite{Chen2016c, Chen2016a}. 
In addition, convex optimization applied to the Hopf-Lax formula allows real-time computation of the HJ PDE solution at any desired state and time instant when the system dynamics are linear \cite{darbon2016algorithms, Chow2017}.

Besides HJ reachability, alternative approaches to verification exist. 
In particular, satisfaction of properties such as safety, liveness, and fairness in computer software and in discrete-time dynamical systems can be verified by checking whether runs of a transition system, or words of a finite automaton satisfy certain desired properties \cite{Baier2008, Belta2017}.
These properties may be specified by a variety of logical formalisms such as linear temporal logic. For specifications of properties of interest in autonomous robots, richer formalisms have been proposed. For example, propositional temporal logic over the reals \cite{Reynolds2001,Fainekos2009} allows specification of properties such as time in terms of real numbers, and chance-constrained temporal logic \cite{Jha2017} allows specification of requirements in the presence of uncertainty. 
Besides autonomous cars and robots, verification approaches based on discrete models have also been successfully used in the context of intelligent transportation systems \cite{Coogan2017} and human-automation interaction \cite{Bolton2013}. 

For continuous and hybrid systems, safety properties can be verified by checking whether the forward reachable set or an over-approximation of it intersects with a set of undesirable states, akin to checking runs of transition systems.
Numerous tools such as SpaceEx \cite{Frehse2011}, Flow* \cite{Chen2013}, CORA \cite{Althoff2015}, C2E2 \cite{Duggirala2015, Fan2016}, and dReach \cite{Kong2015a} have been developed for this purpose; the authors in \cite{Duggirala2016} present a tutorial on combining different tools for hybrid systems verification.
In addition, methods that utilize semidefinite programming to search for Lyapunov functions can be used to verify safety \cite{Parrilo2000, Tedrake2010}. 
This is done, for example, by constructing barrier certificates \cite{Barry2012} or funnels \cite{Majumdar2013,Majumdar2017} with Lyapunov properties. 

Outside of the realm of checking whether the set of possible future states of a system includes undesirable states, safety can also be verified by starting from known unsafe conditions and computing backward reachable sets, which the system should avoid. 
In general, the challenges facing verification methods include computational tractability, generality of system dynamics, existence of control and disturbance variables, and representation of sets \cite{Barron90, Mitchell05, Bokanowski11, Fisac15}.
HJ reachability can be distinguished from other methods because it is applicable to general nonlinear systems, easily handles control and disturbance variables, and is able to represent sets of arbitrary shapes. However, this flexibility comes with the cost of computational complexity. 
Other backward reachability methods make other trade-offs. For example, \cite{Frehse2011, Kurzhanski00, Kurzhanski02, Maidens13} present scalable methods for affine systems that rely on polytopic or ellipsoidal representation of sets, while the methods presented in \cite{Majumdar13, Dreossi16, henrion2014convex} are well-suited to systems with polynomial dynamics.

The goal of this tutorial is four-fold. First, we aim to provide a formal and self-contained introduction to reachability theory. 
Second, we familiarize the readers with some of the available tools for the computation of reachable sets. 
Third, we provide an overview of the recent developments in reachability theory that help overcome the curse of dimensionality. 
Finally, we illustrate some of the recent applications of reachability theory in the verification of safety-critical systems.

\section{Backward Reachable Set (BRS)\label{sec:brs}}
\begin{figure}[t!]
  \centering
  \includegraphics[width=0.95\columnwidth]{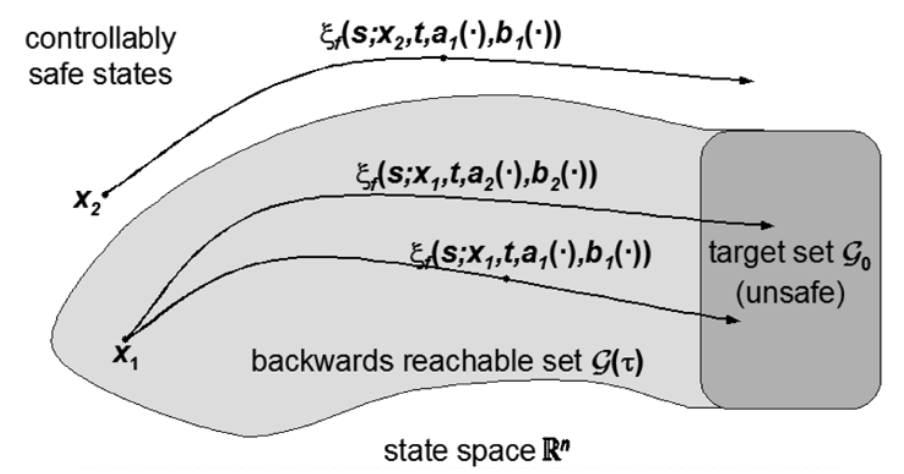}
  \caption{ Target set and backward reachable set. Several trajectories are shown starting at the same time $t$ but from different states $\state$ and subject to different input signals $\ctrl(\cdot)$ and $\dstb(\cdot)$. Input signal $\ctrl(\cdot)$ is chosen to drive the trajectory away from the target set, while input signal $\dstb(\cdot)$ is chosen to drive the trajectory toward the target. Figure taken from \cite{Mitchell05}.}
  \label{fig:brs}
\end{figure}

In reachability theory, we are often interested in computing the \textit{backward reachable set} of a dynamical system. This is the set of states such that the trajectories that start from this set can reach some given target set (see Figure \ref{fig:brs}).
If the target set consists of those states that are known to be unsafe, then the BRS contains states which are potentially unsafe and should therefore be avoided. 
As an example, consider collision avoidance protocols for two aircraft in En-Route airspace. 
The target set would contain those states that are already ``in loss of separation," such as those states in which the aircraft are within the five mile horizontal separation distance mandated by the Federal Aviation Administration. 
The backward reachable set contains those states which could lead to a collision, despite the best possible control actions. 
We typically formulate such safety-critical scenarios in terms of a two-player game, with Player 1 and Player 2 being control inputs. 
For example, Player 1 could represent one aircraft, Player 2 another, with Player 1's control input being treated as the control input of the joint system, and with Player 2's control input being treated as the disturbance.

Mathematically, let $\state\in \R^n$ be the system state, which evolves according to the ordinary differential equation (ODE)
\begin{equation}
\begin{aligned}
\label{eq:fdyn}
\dot\state(s) = \fdyn(\state(s), \ctrl(s), \dstb(s)), s \in [t, 0], \ctrl(s) \in \cset, \dstb(s) \in \dset,
\end{aligned}
\end{equation}
where $\ctrl(s)$ and $\dstb(s)$ denote the input for Player 1 and Player 2 respectively. 
We assume that the control functions $\ctrl(\cdot)$, $\dstb(\cdot)$ are drawn from the set of measurable functions\footnote{A function $f:X\to Y$ between two measurable spaces $(X,\Sigma_X)$ and $(Y,\Sigma_Y)$ is said to be measurable if the preimage of a measurable set in $Y$ is a measurable set in $X$, that is: $\forall V\in\Sigma_Y, f^{-1}(V)\in\Sigma_X$, with $\Sigma_X,\Sigma_Y$ $\sigma$-algebras on $X$,$Y$.}:
\begin{equation*}
\begin{aligned}
\ctrl(\cdot) \in \cfset(t) = & \{\phi: [t, 0] \rightarrow \cset: \phi(\cdot) \text{ is measurable}\}\\
\dstb(\cdot) \in \dfset(t) = & \{\phi: [t, 0] \rightarrow \dset: \phi(\cdot) \text{ is measurable}\}
\end{aligned}
\end{equation*}
\noindent where $\cset \subset \mathbb{R}^{n_u}$ and $\dset \subset \mathbb{R}^{n_d}$ are compact and $t < 0$. 
The system dynamics, or flow field, $\fdyn: \R^n \times \cset \times \dset \rightarrow \R^n$ is assumed to be uniformly continuous, bounded, and Lipschitz continuous in $\state$ uniformly in\footnote{For the remainder of the tutorial, we will omit the notation $(s)$ from variables such as $\state$ and $\ctrl$ when referring to function values.} $\ctrl$ and $\dstb$. 
Therefore, given $\ctrl(\cdot) \in \cfset$ and $\dstb(\cdot) \in \dfset$, there exists a unique trajectory solving \eqref{eq:fdyn} \cite{EarlA.Coddington1955}. 
We will denote solutions, or trajectories of \eqref{eq:fdyn} starting from state $\state$ at time $t$ under control $\ctrl(\cdot)$ and $\dstb(\cdot)$ as $\traj(s; \state, t, \ctrl(\cdot), \dstb(\cdot)): [t, 0] \rightarrow \R^n$. $\traj$ satisfies \eqref{eq:fdyn} with an initial condition almost everywhere:
\begin{equation}
\label{eq:fdyn_traj}
\begin{aligned}
\frac{d}{ds}\traj(s; \state, t, \ctrl(\cdot), \dstb(\cdot)) &= \fdyn(\traj(s; \state, t, \ctrl(\cdot), \dstb(\cdot)), \ctrl(s), \dstb(s)) \\
\traj(t; \state, t, \ctrl(\cdot), \dstb(\cdot)) &= \state
\end{aligned}
\end{equation}

Intuitively, a BRS represents the set of states $\state\in\R^n$ from which the system can be driven into some set $\targetset \subseteq \R^n$ at the \textit{end} of a time horizon of duration $|t|$. 
We call $\targetset$ the ``target set''. 
We assume that Player 1 will try to steer the system away from the target with her input, and Player 2 will try to steer the system toward the target with her input. 
Consequently, we want to compute the following BRS:  
\begin{equation}
\label{eq:BRS}
\begin{aligned}
\brs(t) = & \{\state: \exists \gamma \in \Gamma(t), \forall \ctrl(\cdot) \in \cfset, \\
& \traj(0; \state, t, \ctrl(\cdot), \gamma[\ctrl](\cdot)) \in \targetset\},
\end{aligned}
\end{equation}
where $\Gamma(\cdot)$ in \eqref{eq:BRS} denotes the feasible set of strategies for Player 2. 

The computation of the BRS in \eqref{eq:BRS} requires solving a differential game between Player 1 and Player 2 (more on this in Section \ref{sec:diff_games}).
In a differential game setting, it is important to address what information the players know about each other's decisions which directly affects their strategies, and consequently, the outcome of the game. In reachability problems, we assume that the Player 2 uses only non-anticipative strategies $\Gamma(\cdot)$ \cite{Mitchell05}, defined as follows:
\begin{equation} \label{eqn:nonantistrats}
\begin{aligned}
\gamma \in \Gamma(t) & := \{\mathcal{N}: \cfset(t) \rightarrow \dfset(t):  \ctrl(r) = \hat{\ctrl}(r) \text{ a. e. } r\in[t,s] \\
& \Rightarrow \mathcal{N}[\ctrl](r) = \mathcal{N}[\hat{\ctrl}](r) \text{ a. e. } r\in[t,s]\}
\end{aligned}
\end{equation}
That is, Player 2 cannot respond differently to two~Player 1 controls until they become different. Yet, in this setting, Player 2 has the advantage of factoring in Player 1's choice of input at every instant $t$ and adapting its own accordingly. Thus, Player 2 has an \textit{instantaneous informational advantage}, which allows us to establish safety guarantees under the worst-case scenarios. One particular class of problems in which the notion of non-anticipative strategies is applicable is robust control problems, in which one wants to obtain the robust control (Player 1) with respect to the worst-case disturbance (Player 2), which can then be modeled as an adversary with the instantaneous informational advantage (not because this disturbance is in fact reacting to the controller's input, but rather, because out of all possible disturbances there will be one that will happen to be the worst possible given the chosen control).

The differential game that must be solved in order to compute the BRS in \eqref{eq:BRS} is a ``game of kind" rather than a ``game of degree", i.e., games in which the outcome is determined by \textit{whether or not} the state of the system reaches a given configuration under specified constraints at any time within the duration of the game. The good news is that an approach known as the \textit{level set method} can transform these games of kind into games of degree in an analytically sound and computationally tractable way. We first provide a brief overview of the theory of differential games and then explain how the problem of computing a BRS can be transformed into a differential game of degree using level set methods.

\section{Two-person Zero-sum Differential Games \label{sec:diff_games}}
In many relevant differential game problems, the goal is to optimize a cost function of the final state and some running cost or reward accumulated over system trajectories. The system is steered towards this final state after a finite time horizon. Formally, let $\cost_t(\state, \ctrl(\cdot), \dstb(\cdot))$ denote the cost accumulated during horizon $[t,0]$ when Player 1 and Player 2 play control $\ctrl(\cdot)$ and $\dstb(\cdot)$, respectively. $\cost_t(\cdot)$ can be expressed as
\begin{equation} \label{eqn:cost_fn}
\cost_t(\state, \ctrl(\cdot), \dstb(\cdot)) = \int_{t}^{0} c(\state(s), \ctrl(s), \dstb(s), s)ds + q(\state(0))
\end{equation}

In the zero-sum setting, Player 1 will attempt to maximize this outcome, while the Player 2 will aim to minimize it, subject to the system dynamics in \eqref{eq:fdyn}. 
Under the non-anticipative strategy assumption, we can readily define the so-called \textit{lower value}\footnote{Note that, in general, one needs to define both the upper and lower values of the game, but for the scenarios that we are interested in, the lower value will suffice.} of the game as
\begin{equation} \label{eqn:val_fn}
\valfunc(t, \state) = \inf_{\gamma \in \Gamma(t)} \sup_{\ctrl(\cdot) \in \cfset} \cost_t(\state, \ctrl(\cdot), \gamma[a](\cdot)),
\end{equation}
where $\Gamma(\cdot)$ is defined in \eqref{eqn:nonantistrats}.

Using the principle of dynamic programming, it can be shown that the value function $\valfunc(t, \state)$ in \eqref{eqn:val_fn} is the viscosity solution \cite{evans1983differential} of the following Hamilton-Jacobi Isaacs (HJI) PDE:
\begin{equation} \label{eqn:HJI}
D_t \valfunc(t, \state) + \ham(t, \state, \nabla \valfunc(t, \state)) = 0,\quad \valfunc(0, \state) = q(\state),
\end{equation} 
where $\ham(t, \state, \nabla \valfunc(t, \state))$ is called the Hamiltonian and is given by
\begin{equation} \label{eqn:hamil_diffGame}
\ham(t, \state, \costate) = \max_{\ctrl \in\cset} \min_{\dstb \in\dset}  c(\state, \ctrl, \dstb, t) + \costate \cdot \fdyn(\state, \ctrl, \dstb).
\end{equation} 
$\costate$ in \eqref{eqn:hamil_diffGame} denotes $\nabla \valfunc(t, \state)$ and is called the \textit{costate}. Given the value function, the optimal control for Player 1 can be obtained as:
\begin{equation}
\label{eq:OptCtrl_diffgame}
\ctrl^*(t, \state) = \arg \max_{\ctrl \in\cset} \min_{\dstb \in\dset}  c(\state, \ctrl, \dstb, t) + \costate \cdot \fdyn(\state, \ctrl, \dstb).
\end{equation} 
The optimal control for Player 2 can be similarly obtained. A more detailed discussion of this material can be found in \cite{evans1983differential}.

\section{The Level Set Approach: From Games of Kind to Games of Degree \label{sec:level_set}}
We are now ready to solve the original intended problem of this tutorial: the computation of BRS. In Section \ref{sec:diff_games}, we discussed how the differential games of degree can be solved using an HJ PDE. The computation of the BRS, however, is a differential game of kind where the outcome is Boolean: the system either reaches the target set or not. It turns out that we can ``encode" this Boolean outcome through a quantitative value function: for example, if we consider $\cost_t(\cdot)$ as the distance between the system state and the target region at the terminal state of the system, it is easy to determine whether the system reached the target by comparing this distance to some threshold value (simply 0 in this case). This allows us to find the solution to a game of kind by posing an auxiliary game of degree whose solution encodes that of the original problem: this is, in essence, the level set approach.

In particular, one can always find a Lipschitz function $\ic(\state)$ such that $\targetset$ (the target set) is equal to the zero sublevel set of $\ic$, that is, $\state \in \targetset \Leftrightarrow \ic(\state) \le 0$. The Lipschitz function $\ic$ can always be found, since one can always choose the signed distance to the respective sets. If we define the cost function to be
\begin{equation} \label{eqn:cost_BRS}
\cost_t(\state, \ctrl(\cdot), \dstb(\cdot)) = \ic(\state(0)),
\end{equation}
then the system reaches the target set under controls $\ctrl$ and $\dstb$ if and only if $\cost_t(\state, \ctrl(\cdot), \dstb(\cdot)) \le 0$. Since Player 2 wants to drive the system to the target, it wants to minimize the cost in \eqref{eqn:cost_BRS}, and Player 1 wants to maximize this cost. We can now compute the value function $\valfunc(t, \state)$ for this differential game in a similar fashion to Section \ref{sec:diff_games}. Consequently, the BRS can be obtained as 
\begin{equation}
\brs(t) = \{\state: \valfunc(t, \state) \le 0\},
\end{equation}
where $\valfunc(t, \state)$ satisfies the following HJI PDE:
\begin{equation} \label{eqn:brs_PDE}
D_t \valfunc(t, \state) + \ham(t, \state, \costate) = 0,\quad \valfunc(0, \state) = \ic(\state).
\end{equation} 
The Hamiltonian is given by
\begin{equation} \label{eqn:hamil_brs}
\ham(t, \state, \costate) = \max_{\ctrl \in\cset} \min_{\dstb \in\dset} \costate \cdot \fdyn(\state, \ctrl, \dstb).
\end{equation} 

The interpretation of $\brs(t)$ is that if $\state(t) \in \brs(t)$, then Player 2 has a control sequence that will drive the system to the target at time $0$, irrespective of the control of Player 1. If $\state(t) \in \partial\brs(t)$, where $\partial\brs(t)$ denotes the boundary of $\brs(t)$, then Player 1 will \textit{barely} miss the target at time $0$ if it applies the optimal control
\begin{equation}
\label{eq:OptCtrl_brs}
\ctrl^*(t, \state) = \arg \max_{\ctrl \in\cset} \min_{\dstb \in\dset}  \costate \cdot \fdyn(\state, \ctrl, \dstb).
\end{equation} 
Finally, if $\state(t) \in \brs(t)^C$, then Player 1 has a control sequence (given by \eqref{eq:OptCtrl_brs}) that will keep the system out of the target set, irrespective of the control applied by Player 2. In particular, when the target set $\targetset$ represents unsafe/undesired states of the system and Player 2 represents the disturbances in the system, then $\brs(t)$ represents the \textit{effective} unsafe set, i.e., the set of states from which the disturbance can drive the system to the \textit{actual} unsafe set despite the best control efforts. Thus, reachability analysis gives us the safe set (in this case $\brs(t)^C$) as well as a controller (in this case $\ctrl^*(t, \state)$) that will keep the system in the safe set, given that the system starts in the safe set. 

\section{Different flavors of reachability \label{sec:flavors_reach}}
So far, we have presented the computation of BRSs, but reachability analysis is not limited to BRSs. One can compute various other kinds of sets that may be more useful, depending on the verification problem at hand. In this section, we provide a brief overview of some of these sets.

\subsection{Forward vs. Backward Reachable Set \label{sec:frs}}
In some cases, we might be interested in computing a forward reachable set (FRS): the set of all states that a system can reach from a given initial set of states after a time duration of $|t|$. Formally, we want to compute the following set:
\begin{equation}
\label{eq:FRS}
\begin{aligned}
\frs(t) = & \{y: \exists \gamma \in \Gamma(t), \forall \ctrl(\cdot) \in \cfset,  \\
& \traj(t; \state, 0, \ctrl(\cdot), \gamma[\ctrl](\cdot)) = y, \state \in \targetset\}, t>0.
\end{aligned}
\end{equation}
Here, $\targetset$ represents the set of initial states of system. $\frs(t)$ is the set of all states that system can reach in a duration of $t$, while Player 1 applies the control to keep the system in $\targetset$ and Player 2 applies the control to drive the system out of $\targetset$. The FRS can be computed in a similar fashion as the BRS. The only difference is that an initial value HJ PDE needs to be solved instead of a final value PDE, which can always be converted into an equivalent final value PDE by change of variables \cite{evans2010partial}. More details on the computation of FRS and some of their concrete applications can be found in \cite{Chen2016d, mitchell2007comparing}.

\subsection{Reachable Sets vs. Tubes \label{sec:brs_brt}}
Another important aspect in reachability is that of reachable tubes. The reachable set is the set of states from which the system can reach a target at \textit{exactly} time $0$. Perhaps a more useful notion is to compute the set of states from which the system can reach a target \textit{within} a duration of $|t|$. For example, for safety analysis, we are interested in verifying if a disturbance can drive the system to the unsafe states \textit{ever} within a horizon, and not just at the end of the horizon. This notion is captured by reachable tubes. Here, we present the formal definition of backward reachable tube (BRT), but forward reachable tube (FRT) can be similarly defined:
\begin{equation}
\label{eq:BRT}
\begin{aligned}
\brs(t) = & \{\state: \exists \gamma \in \Gamma(t), \forall \ctrl(\cdot) \in \cfset, \\
& \exists s \in [t, 0], \traj(s; \state, t, \ctrl(\cdot), \gamma[\ctrl](\cdot)) \in \targetset\}.
\end{aligned}
\end{equation}
Once again, the BRT can be computed by solving a final value PDE similar to that in \eqref{eqn:brs_PDE} \cite{Mitchell05, lygeros2004reachability}.

\subsection{Roles of the Control and Disturbance\label{sec:minmax}}
Depending on the role of Player 1 and Player 2, we may need to use different max-min combinations. As a rule of thumb, whenever the existence of a control (``$\exists \ctrl$") is sought, the optimization is a minimum over the set of controls in the corresponding Hamiltonian. Whenever a set/tube characterizes the behavior of the system for all controls (``$\forall \ctrl$"), the optimization is a maximum. For example, for the BRS in \eqref{eq:BRS}, we sought the \textit{existence} of a Player 2 controller \textit{for all} Player 1 controls, so we used minimum for Player 2 and maximum for Player 1 in the Hamiltonian (see \eqref{eqn:hamil_brs}). When the target set represents the set of the desired states that we want the system to reach and Player 2's control represents the disturbance, then we are interested in verifying if there exists a control of Player 1 such that the system reaches its target despite the worst-case disturbance. In this case, we should use maximum for Player 2's control and minimum for Player 1's control in the corresponding Hamiltonian.

\subsection{Presence of State Constraints}
Another interesting problem that arises in verification is the reachability to and from a target set subject to some state constraints; this can be handled efficiently for even time-dependent constraints within the reachability framework \cite{Margellos11,Fisac15}. In general, any combination of the above four variants can be solved using the HJ reachability formulation. Partially, it is this flexibility of the reachability framework that has facilitated its use in various safety-critical applications, some of which we will discuss in this tutorial.

\section{Computational Tools for HJ Reachability\label{sec:code}}
In this section, we will present an overview of two available computational tools that can be used to compute different definitions of reachable sets.
\subsection{The Level Set Toolbox (toolboxLS)}
The level set toolbox (or \textit{toolboxLS}) was developed by Professor Ian Mitchell \cite{Mitchell07b} to solve partial differential equations using level set methods, and is the foundation of the HJ reachability code. The toolbox is implemented in MATLAB and is equipped to solve any final-value HJ PDE.  Since different reachable set computations can be ultimately posed as solving a final-value HJ PDE (see Sections \ref{sec:level_set} and \ref{sec:flavors_reach}), the level set toolbox is fully equipped to compute various types of reachable sets. Information on how to install and use toolboxLS can be found here: \textit{http://www.cs.ubc.ca/$\sim$mitchell/ToolboxLS}. This toolbox can be further augmented by the Hamilton-Jacobi optimal control toolbox (or \textit{helperOC}). A quick-start guide to using toolboxLS and helperOC is presented in the Appendix and is also available at:\textit{ http://www.github.com/HJReachability/helperOC}.

\subsection{The Berkeley Efficient API in C++ for Level Set methods (BEACLS) Toolbox}
The Berkeley Efficient API in C++ for Level Set methods (\textit{BEACLS}) Toolbox was developed by Ken Tanabe. This toolbox implements the functions from helperOC and toolboxLS in C++ for fast computation of reachability analyses. The library also uses GPUs for parallelizing different computations in the level set toolbox. The installation instructions and user guide can be found at: \textit{http://www.github.com/HJReachability/beacls}.  This GPU library has been used for large-scale multi-vehicle reachability problems, such as safe path planning (see Section \ref{sec:spp}).

\section{Current Research in HJ Reachability Theory\label{sec:currentWork}}
Recently there have been several advances in HJ reachability theory and applications. Research on restructuring dynamics, new formulations for analysis, and the addition of learning techniques provided HJ reachability with a broadened and deeper span of feasible applications. These advances are used in safety-critical applications to provide safety guarantees, liveness properties, and optimal controllers.

\subsection{System Decomposition Techniques for Nonlinear Systems \label{sec:Decomp}}
\begin{figure}[!htb]
  \centering
  \begin{subfigure}[b]{0.49\columnwidth}
    \includegraphics[trim={12cm 0 3cm 0},clip,width=\columnwidth]{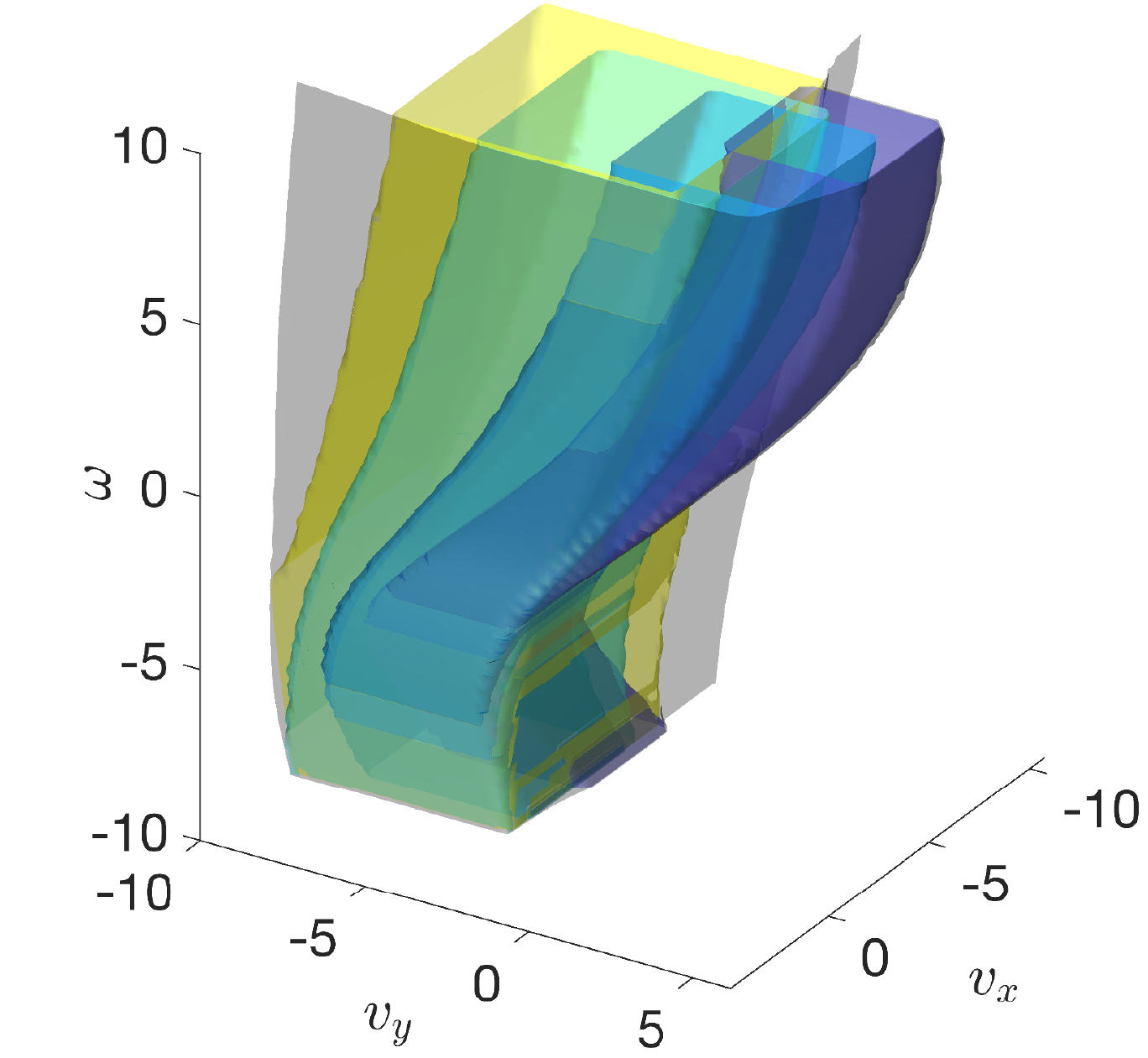}
    \subcaption{BRS \& BRT for a 6D\\ quadrotor avoiding an obstacle.}
    \label{fig:Quad6D}
  \end{subfigure}%
  \begin{subfigure}[b]{0.49\columnwidth}
    \includegraphics[trim={10cm 0 0 2cm},clip,width=\columnwidth]{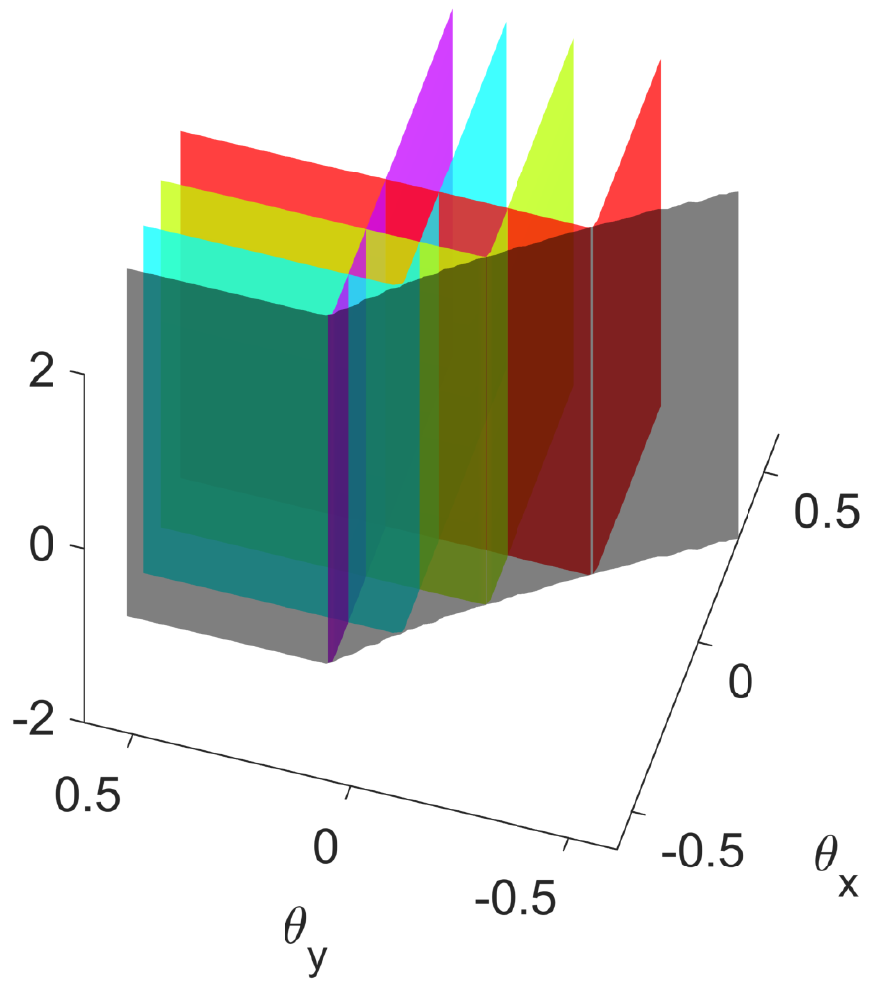}
    \subcaption{Reachable set and tube for a 10D quadrotor reaching a target.}
    \label{fig:Quad10D}
  \end{subfigure}%
  \caption{Decomposition results for nonlinear systems. Figures taken from \cite{Chen2016c}.}
  \label{fig:decomp_nonlinear}
\end{figure}

Decomposition methods address the exponentially scaling computational complexity of previous approaches for solving HJ reachability problems, which makes application to high-dimensional systems intractable. In \cite{Chen2016b, Chen2016c} a new technique is proposed that decomposes the dynamics of a general class of nonlinear systems into subsystems which may be coupled through common states, controls, and disturbances. Despite this coupling, BRSs and BRTs can be computed efficiently and exactly using this technique without the need for linearizing dynamics or approximating sets as polytopes. Computations of BRSs and BRTs now become orders of magnitude faster, and for the first time BRSs and BRTs for many high-dimensional nonlinear control systems can be exactly computed. In situations where the exact solution cannot be computed, this method can obtain slightly conservative results. The paper demonstrates this theory by numerically computing BRSs and BRTs for several systems, including the 6D Acrobatic Quadrotor and the 10D near-Hover Quadrotor. Reachable sets computed using the decomposition process are illustrated in Figure \ref{fig:decomp_nonlinear}, with details in \cite{Chen2016b, Chen2016c}.

In more general settings, approximate decomposition of nonlinear systems can be achieved by treating key states as disturbances, as in \cite{Chen2016a, Mitchell03}. These methods are able to maintain a direction of conservatism in order to provide guarantees on system performance and safety by either computing overapproximations or underapproximations of reachable sets and tubes. In \cite{Chen2016a}, the authors also propose a way to trade off conservatism of the solution with computational cost.

\begin{figure}[!htb]
  \centering
  \begin{subfigure}[b]{0.45\columnwidth}
    \includegraphics[trim={0 0 0 0},clip,width=\columnwidth]{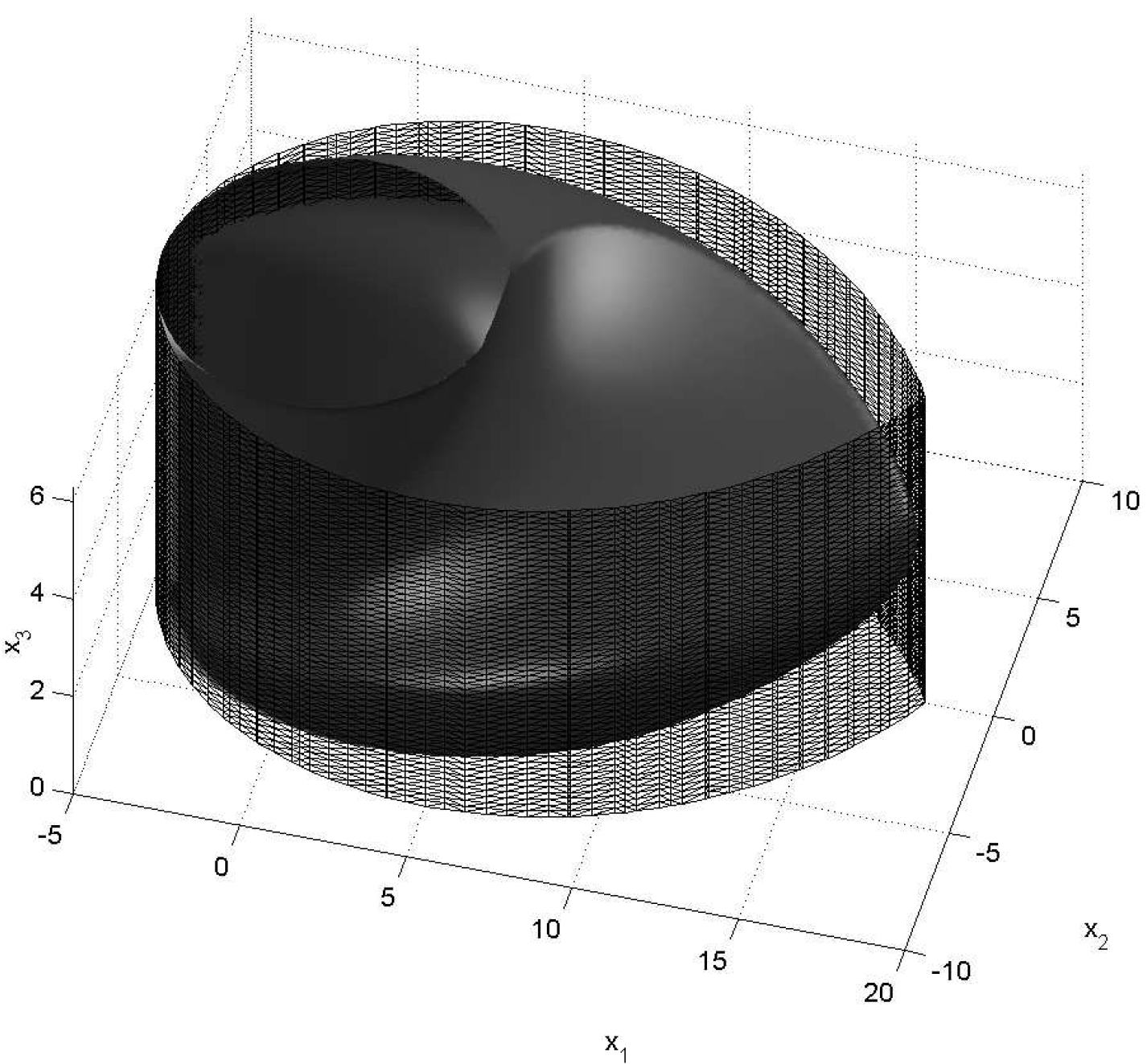}
    \subcaption{Projection-based approximation of a reachable tube. Figure taken from \cite{Mitchell03}.}
  \end{subfigure}~~%
  \begin{subfigure}[b]{0.45\columnwidth}
    \includegraphics[trim={0 0 21cm 3cm},clip,width=\columnwidth]{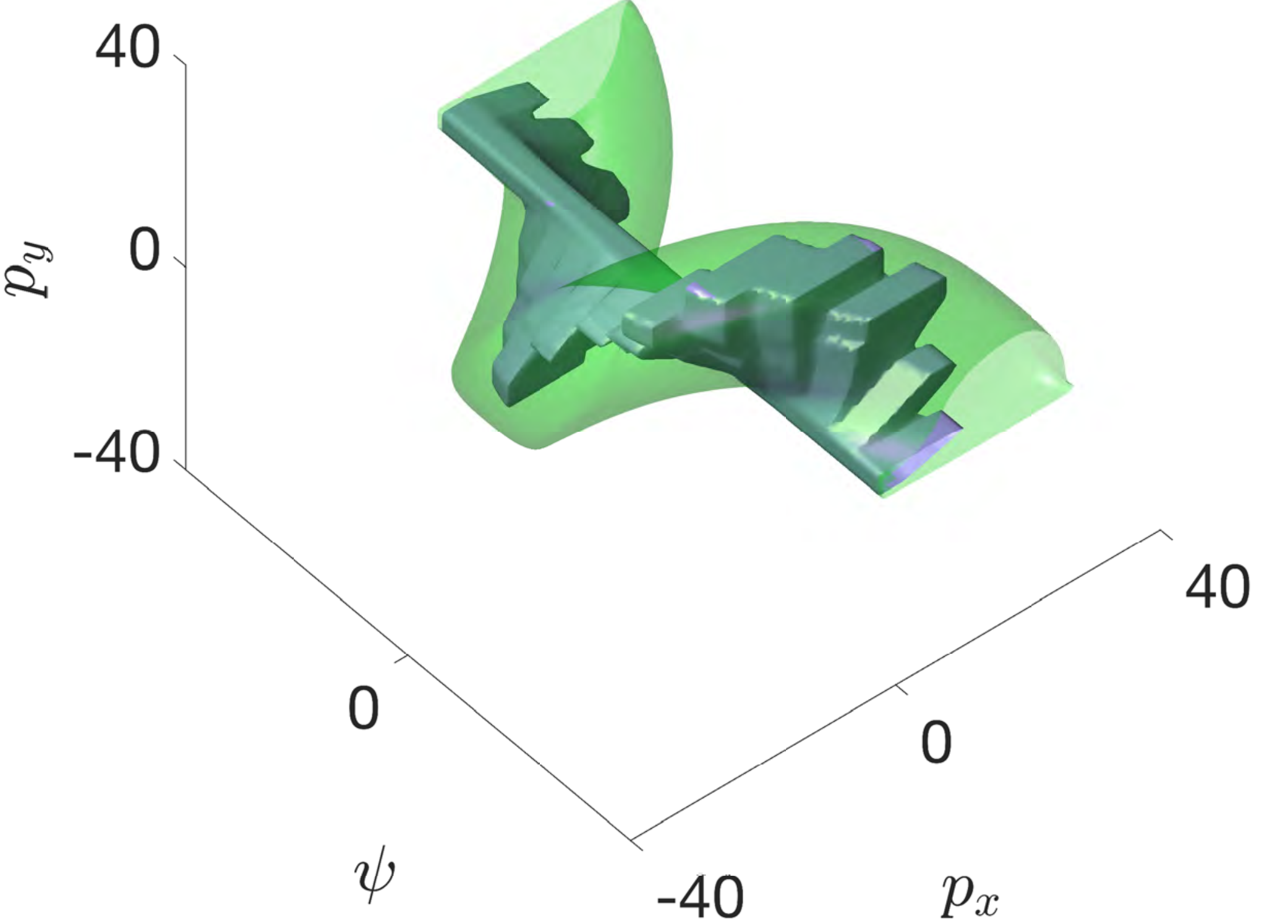}
    \subcaption{Decoupling disturbance-based approximation of a reachable set. Figure taken from \cite{Chen2016a}.}
  \end{subfigure}%
  \caption{Approximate decomposition results for nonlinear systems.}
  \label{fig:decompApprox}
  \vspace{-.3in}
\end{figure}
\subsection{System Decomposition Techniques for Linear Time-Invariant Systems}
\begin{figure}[!htb]
  \centering
  \begin{subfigure}{0.45\columnwidth}
    \includegraphics[width=\columnwidth]{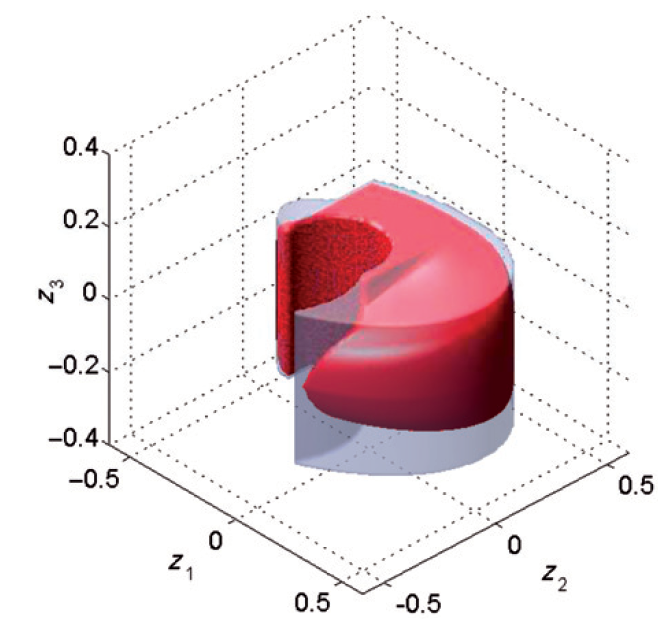}
    \subcaption{Overapproximation (translucent) of a reachable set (solid). Figure taken from \cite{Kaynama2011}.}
    \label{fig:sf_d6sep0}
  \end{subfigure}~~%
  \begin{subfigure}{0.45\columnwidth}
    \includegraphics[width=\columnwidth]{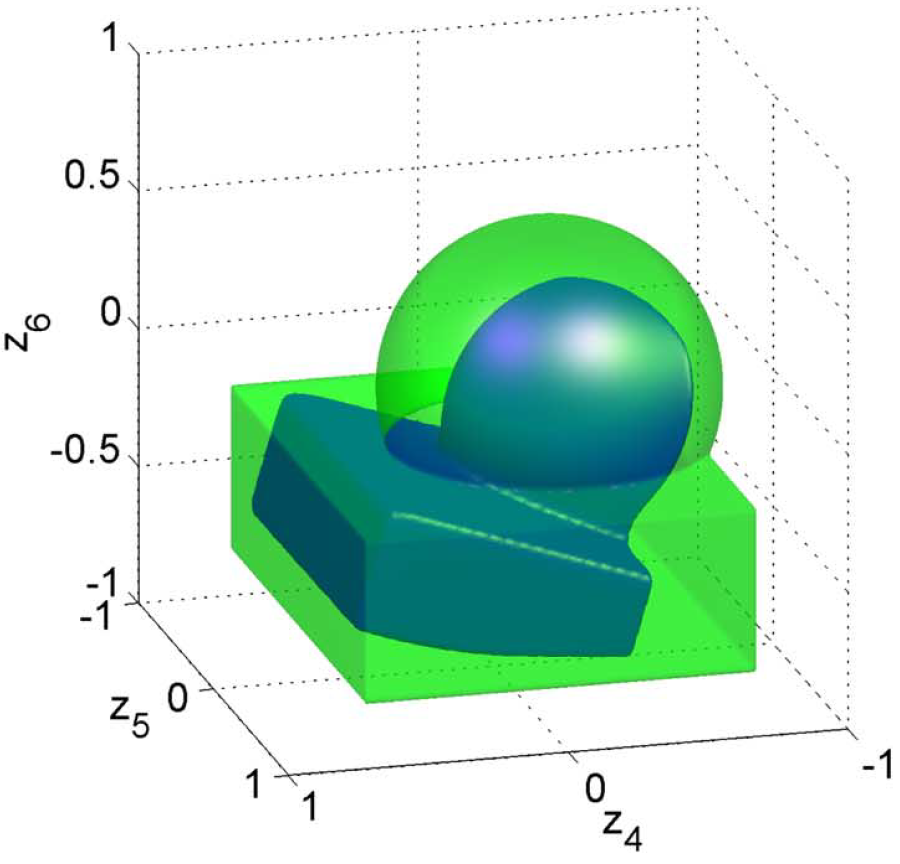}
    \subcaption{Constraint set (translucent) and the approximate viability kernel (solid). Figure taken from \cite{Kaynama2013}.}
    \label{fig:sf_d11sep0}
  \end{subfigure}%
  \caption{Decomposition results for linear time-invariant systems.}
  \label{fig:decomp_LTI}
  \vspace{-.1in}
\end{figure}

In the linear time-invariant case, many non-HJ-based computation techniques have been developed for approximating reachable sets. In the area of HJ reachability, specific decomposition techniques also exist, and provide a substantial reduction in computational burden with a small degree of conservatism. In \cite{Kaynama2011}, the authors proposed a Schur-based decomposition technique for computing reachable sets and synthesizing safety-preserving controllers. Subsystems are analyzed separately, and reachable sets of subsystems are back-projected and intersected to construct an overapproximation of the reachable set, so that safety can still be guaranteed. In \cite{Kaynama2013}, a similar approach based on a modified Riccati transformation is used. Here, decentralized computations are done in transformed coordinates of subspaces. The computation results are combined to obtain an approximation of the viability kernel, which is the complement of the reachable set. Figure \ref{fig:decomp_LTI} shows the conservative approximations obtained from these decomposition techniques.

\subsection{Fast and Safe Tracking for Motion Planning\label{sec:FaSTrack}}
\begin{figure}[h]
	\centering
	\includegraphics[width=.8\columnwidth]{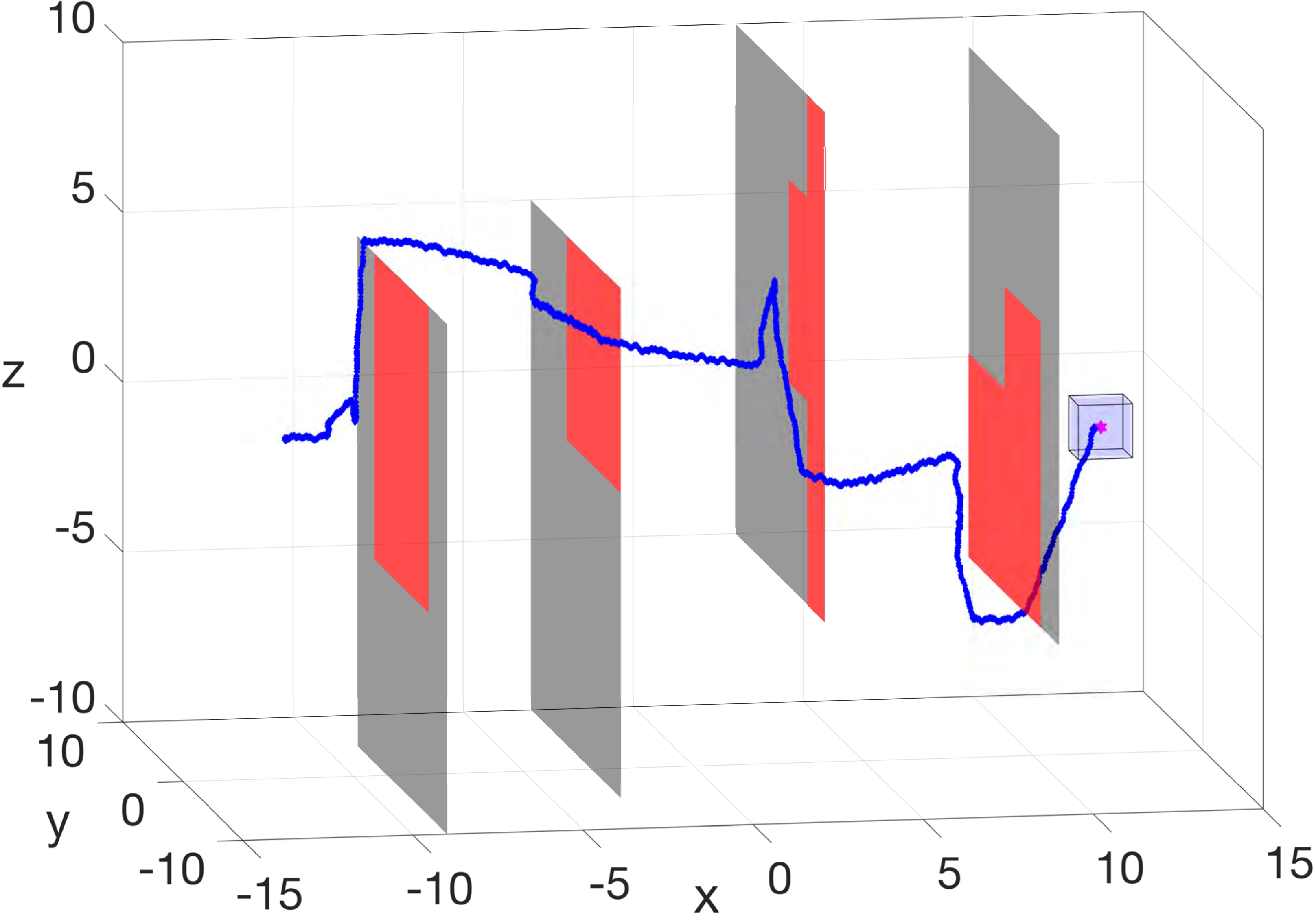}
	\caption{Real-time safe planning using FaSTrack. Figure obtained from \cite{Herbert2017}.}
	\label{fig:fastrack}
	\vspace{-.2in}
\end{figure}

Fast and safe navigation of dynamical systems through a priori unknown cluttered environments is vital to many applications of autonomous systems. However, trajectory planning for autonomous systems is computationally intensive, often requiring simplified dynamics that sacrifice safety and dynamic feasibility in order to plan efficiently. Conversely, safe trajectories can be computed using more sophisticated dynamic models, but this is typically too slow to be used for real-time planning. In \cite{Herbert2017}, a new algorithm is developed called FaSTrack: Fast and Safe Tracking. A path or trajectory planner using simplified dynamics to plan quickly can be incorporated into the FaSTrack framework, which provides a safety controller for the vehicle along with a guaranteed tracking error bound. By formulating a differential game and leveraging HJ reachability's flexibility with respect to nonlinear system dynamics, this tracking error bound is computed in the error coordinates, which evolve according to the error dynamics, and captures all possible deviations due to dynamic infeasibility of the planned path and external disturbances. Note that FaSTrack is modular and can be used with other path or trajectory planners. This framework is demonstrated using a 10D nonlinear quadrotor model tracking a 3D path obtained from an RRT planner, shown in Figure \ref{fig:fastrack}.
\subsection{HJ Reachability for Safe Learning-Based Control\label{sec:JaimeKene}}
The proven efficacy of learning-based control schemes strongly motivates their application to  robotic systems operating in the physical world. However, guaranteeing correct operation during the learning process is currently an unresolved issue, which is of vital importance in safety-critical systems.
\begin{figure}[H]
  \centering
  \includegraphics[width=1\columnwidth]{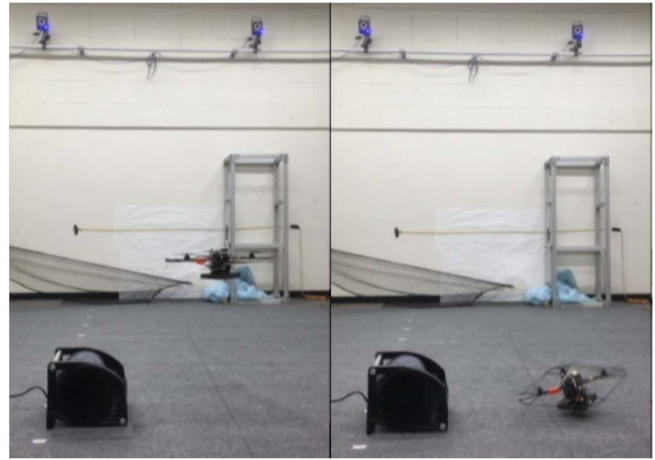}
  \caption{A Hummingbird UAV is able to successfully reject disturbances using online learning, and fails to do so without learning. Figure obtained from \cite{Fisac2017}.}
  \label{fig:safeLearning}
\end{figure}
\vspace{-.1in}

In \cite{Akametalu2014, Fisac2017}, a general safety framework is proposed based on HJ reachability methods that can work in conjunction with an arbitrary learning algorithm. The method exploits approximate knowledge of the system dynamics to guarantee constraint satisfaction while minimally interfering with the learning process. The authors further introduce a Bayesian mechanism that refines the safety analysis as the system acquires new evidence, reducing initial conservativeness when appropriate while strengthening guarantees through real-time validation. The result is a least-restrictive, safety-preserving control law that intervenes only when (a) the computed safety guarantees require it, or (b) confidence in the computed guarantees decays in light of new observations.

The authors provide safety guarantees combining probabilistic and worst-case analysis and demonstrate the proposed framework experimentally on a quadrotor vehicle. Even though safety analysis is based on a simple point-mass model, the quadrotor is able to successfully run policy-gradient reinforcement learning without crashing, and safely retracts away from a strong external disturbance introduced during one of the experiments, as shown in Figure \ref{fig:safeLearning}.
\subsection{HJ Reachability Analysis using Neural Networks \label{sec:Vicenc}}
\begin{figure}[H]
  \centering
  \begin{subfigure}{0.48\columnwidth}
    \includegraphics[width=\columnwidth]{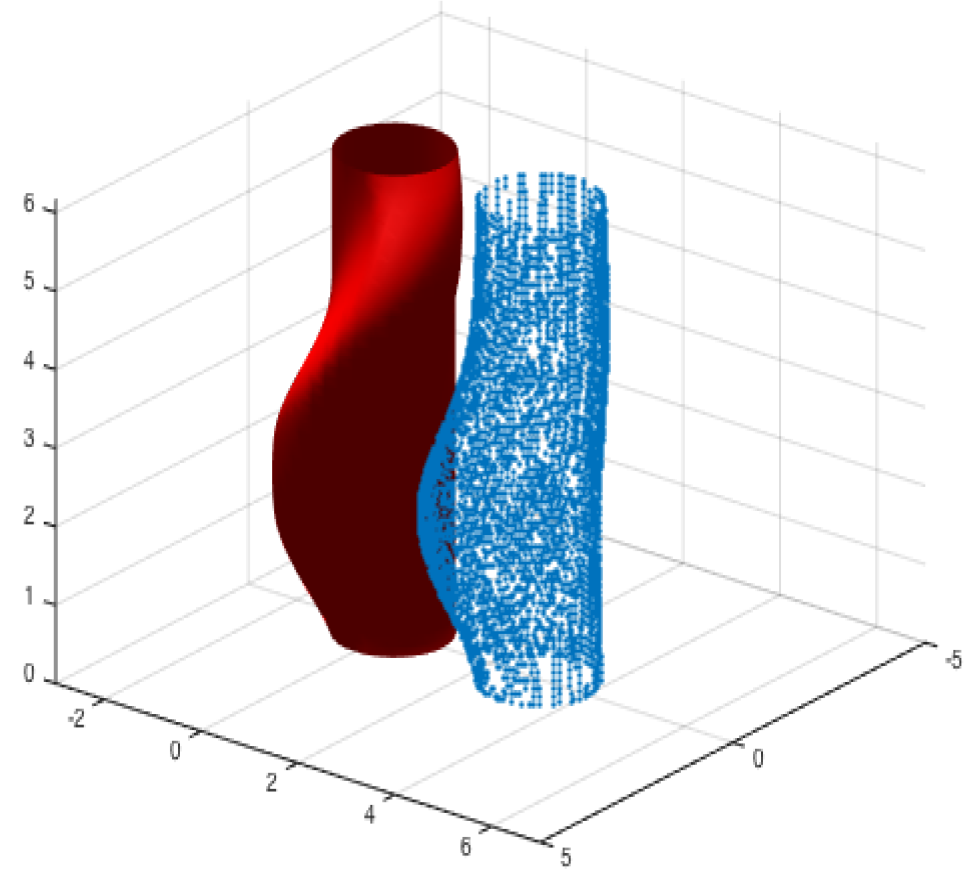}
    \subcaption{Approximation a reachable set (red) using a neural network (point cloud). Figure taken from \cite{Royo2016}.}
  \end{subfigure}~~%
  \begin{subfigure}{0.48\columnwidth}
    \includegraphics[width=\columnwidth]{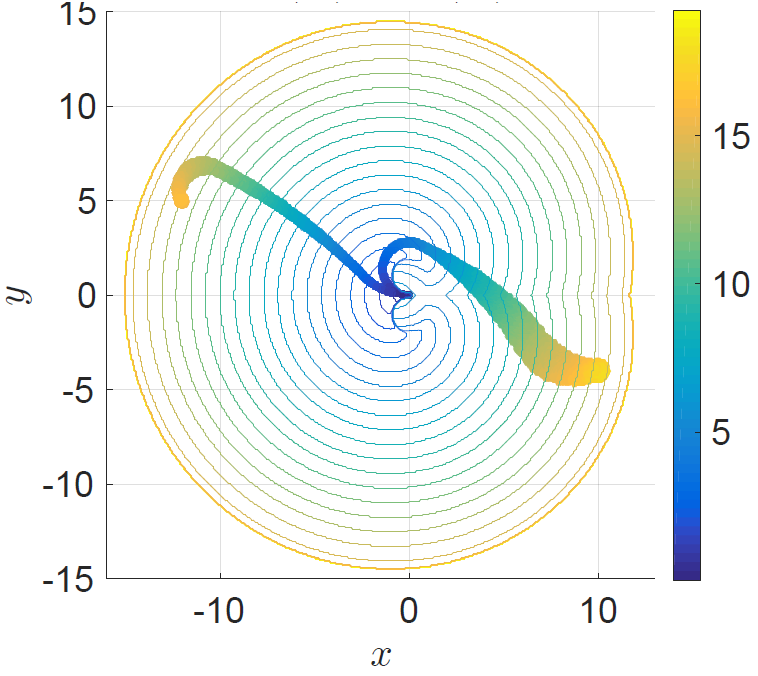}
    \subcaption{Overapproximation of a value function (contours) using a neural network (color gradient) in small regions of the state space. Figure taken from \cite{Jiang2016}.}
  \end{subfigure}%
  \caption{Neural network-based approximations of value functions representing reachable sets.}
  \label{fig:NN}
\end{figure}

Many of the recent breakthroughs in machine learning and AI have been possible thanks in part to the use of powerful function approximators, and in particular (deep) neural networks. In AI, these approximators are used to represent a myriad of complex functions such as Value functions, Q-functions and control policies, which often have high-dimensional data as inputs. In \cite{Niarchos2006,Djeridane2006,Royo2016,Jiang2016}, the authors use these same tools in the context of reachability to approximate solutions of the HJ PDE by implementing and analyzing learning-based algorithms to approximate the solution of certain types of HJ PDEs using neural networks. Some recent results on 2D and 3D systems show that these learning-based algorithms require less memory to run and less memory to store the resulting approximation than traditional gridding-based methods. Further work involves exploring how well these algorithms scale with the number of dimensions in the state space, as well as the types of safety guarantees that can be derived from these types of approximations. In some cases, conservative guarantees for the computed value functions are possible despite the use of neural networks. Figure \ref{fig:NN} shows preliminary results.

\subsection{Generalized Hopf Formula for Linear Systems \label{sec:Osher}}
\begin{figure}[!htb]
  \centering
  \includegraphics[width=0.6\columnwidth]{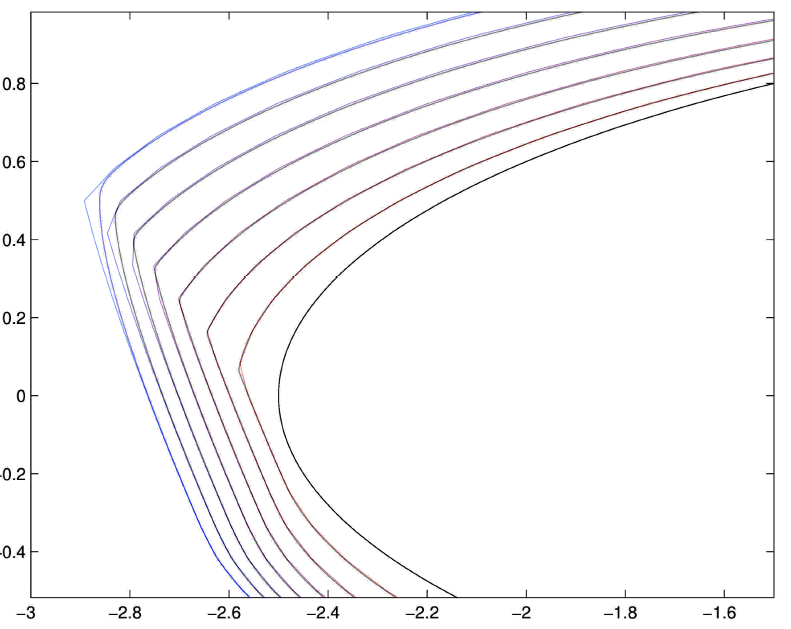}
  \caption{Comparison between HJ PDE solutions obtained using the Hopf formula (colored) and using Lax-Friedrichs finite difference (black-and-white). Figure obtained from \cite{Chow2017}.}
  \label{fig:Hopf}
\end{figure}

In \cite{darbon2016algorithms,Chow2017}, the authors proposed using a generalized Hopf formula for solving HJ PDEs arising from linear systems, which may be time-varying. Obtaining HJ PDE solutions here involves solving the minimization problem in the generalized Hopf formula. This minimization problem can be solved using any optimization algorithm; the authors suggest using coordinate descent with multiple initializations, as well as a numerical quadrature rule for an integral with respect to time. Alternative algorithms such as ADMM can also be used. By reformulating the problem of solving the HJ PDE as an optimization problem, the solution for HJ PDEs can be obtained at any desired points in state space and time, effectively alleviating the exponentially scaling computational complexity in finite difference-based methods. Figure \ref{fig:Hopf} shows the results of this method.

\section{Some Current Applications of HJ Reachability \label{sec:currentApp}}

\subsection{Unmanned Aerial Systems Traffic Management (UTM) using Air Highways}
\begin{figure}
  \centering
  \begin{subfigure}[t]{0.75\columnwidth}
    \includegraphics[width=\columnwidth]{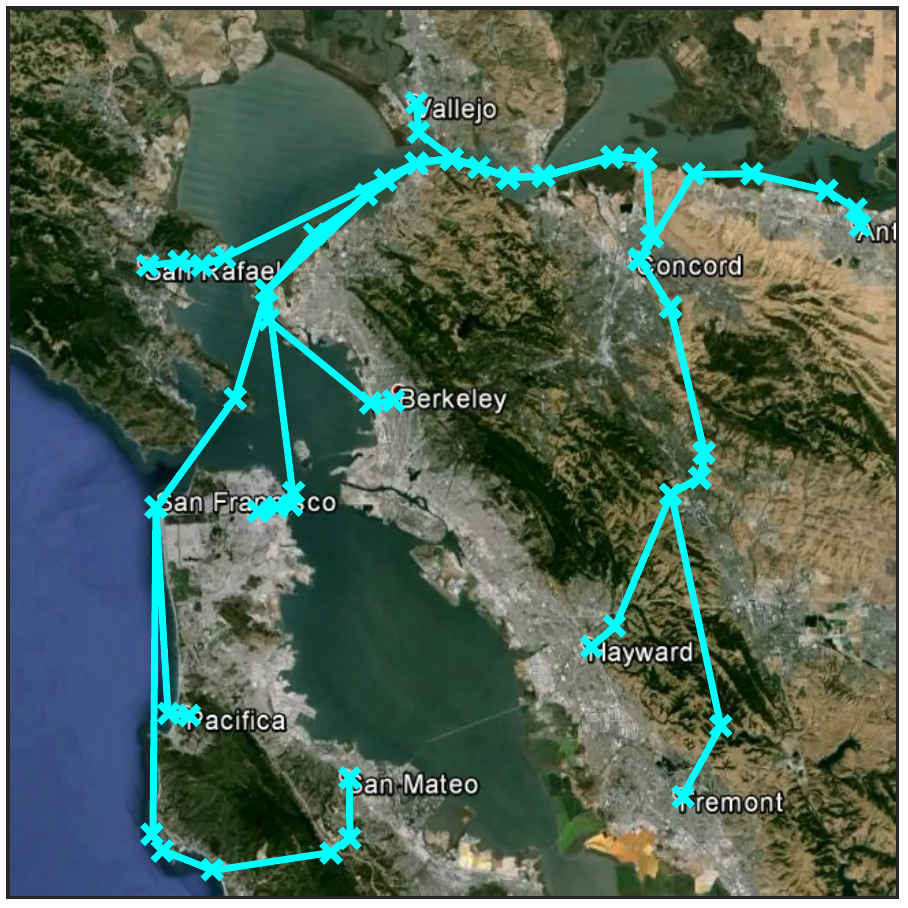}
    \caption{Air highway placement over the San Francisco Bay Area. \label{subfig:airHighway}}
  \end{subfigure}
  \begin{subfigure}[t]{0.9\columnwidth}
    \includegraphics[width=\columnwidth]{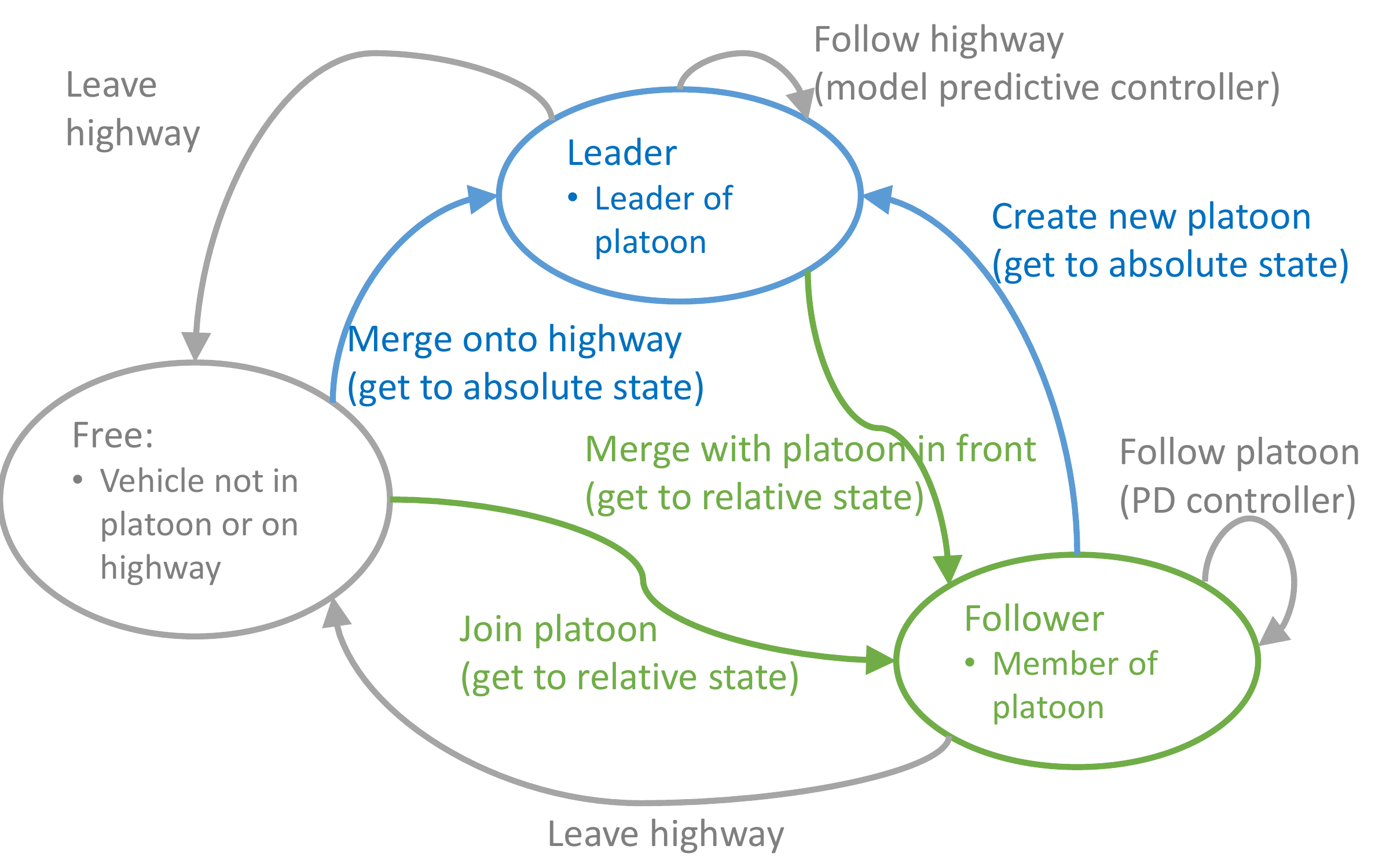}
    \caption{The purple vehicle is joining the platoon while avoiding collisions. \label{subfig:modeControllers}}
  \end{subfigure}
  
  \caption{The air highway and platooning concept for UTM. Figures are taken from \cite{Chen2017}.}
\end{figure}

In collaboration with the National Aeronautics and Space Administration (NASA), HJ reachability has been applied to UTM \cite{Kopardekar16}. In \cite{Chen15b,Chen2017}, the authors proposed an efficient and flexible method for the placement of air highways, which are designated virtual pathways in the airspace. Air highways provide a scalable and intuitive way for monitoring and managing a large number of unmanned aerial vehicles (UAVs) flying in civilian airspace. The proposed method starts with a cost map encoding the desirability of having UAVs fly in different parts of a region, and computes minimum-cost paths connecting origins and destinations. These paths can be updated in real time according to changes in the airspace. Trunks and branches of air highways, similar to ground-based highway systems, naturally emerge from the proposed method. Applying the method to the San Francisco Bay Area, these air highways, which avoid urban areas and airports as much as possible, are shown in Figure \ref{subfig:airHighway}.

To fulfill potential traffic rules on the air highways, a hybrid system model for each UAV is used. On the highway system, a UAV can be in the ``Free'', ``Leader'' or ``Follower'' modes. In this context, HJ reachability is used to ensure the success and safety of mode transitions. For example, the transition from the Free mode to the Leader mode involves using a controller from a maximal backward reachable set to arrive at a prescribed destination on the highway at a prescribed time. The highway and platoon structure greatly reduces the chance of multiple conflicts, enabling the use of pairwise safety analysis. Pairwise safety can be guaranteed using a minimal backward reachable set defined in the relative coordinates of two vehicles. The hybrid systems model is shown in Figure \ref{subfig:modeControllers}. The proposed platooning concept has been implemented in the quadrotor lab at UC Berkeley on Crazyflies 2.0, which is an open source nano quadrotor platform developed by
Bitcraze.

\subsection{Sequential Robust Space-Time Reservations \label{sec:spp}} 
\begin{figure}[]
  \centering
  \begin{subfigure}{0.48\columnwidth}
    \includegraphics[width=\columnwidth]{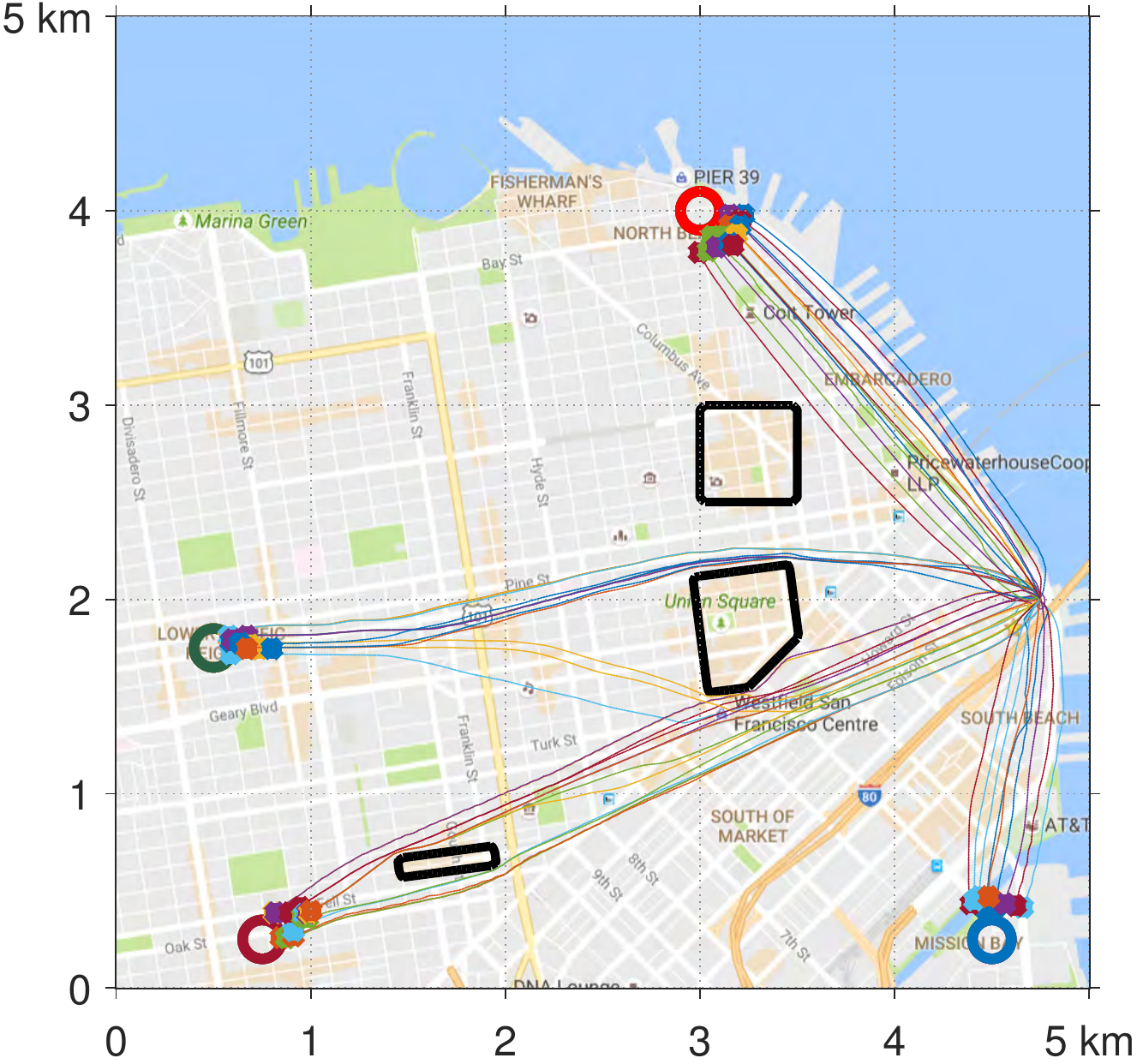}
    \subcaption{6 m/s wind, high UAV density}
    \label{fig:sf_d6sep0}
  \end{subfigure}~%
  \begin{subfigure}{0.48\columnwidth}
    \includegraphics[width=\columnwidth]{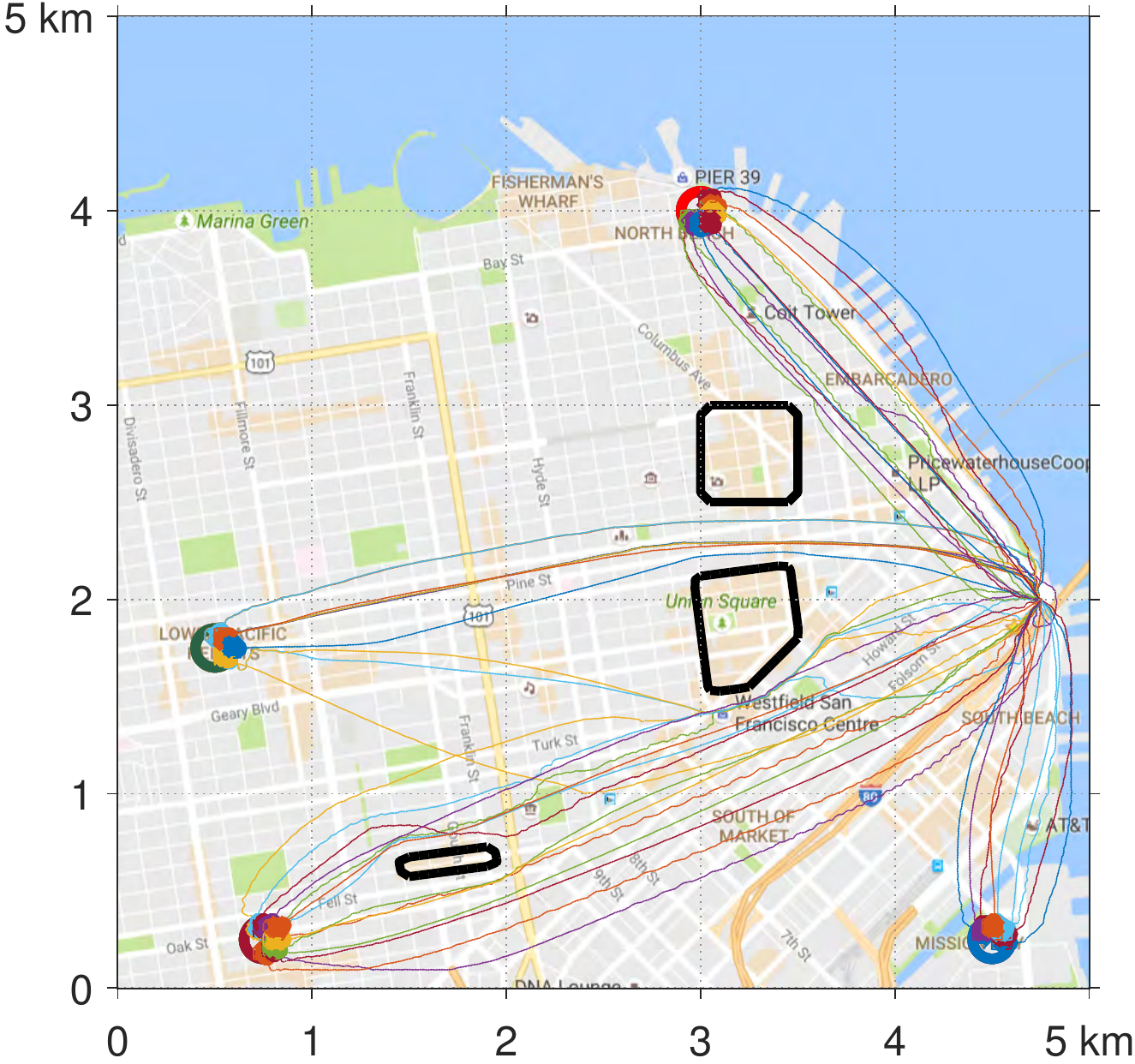}
    \subcaption{11 m/s wind, high UAV density}
    \label{fig:sf_d11sep0}
  \end{subfigure}%
  
  \begin{subfigure}{0.48\columnwidth}
    \includegraphics[width=\columnwidth]{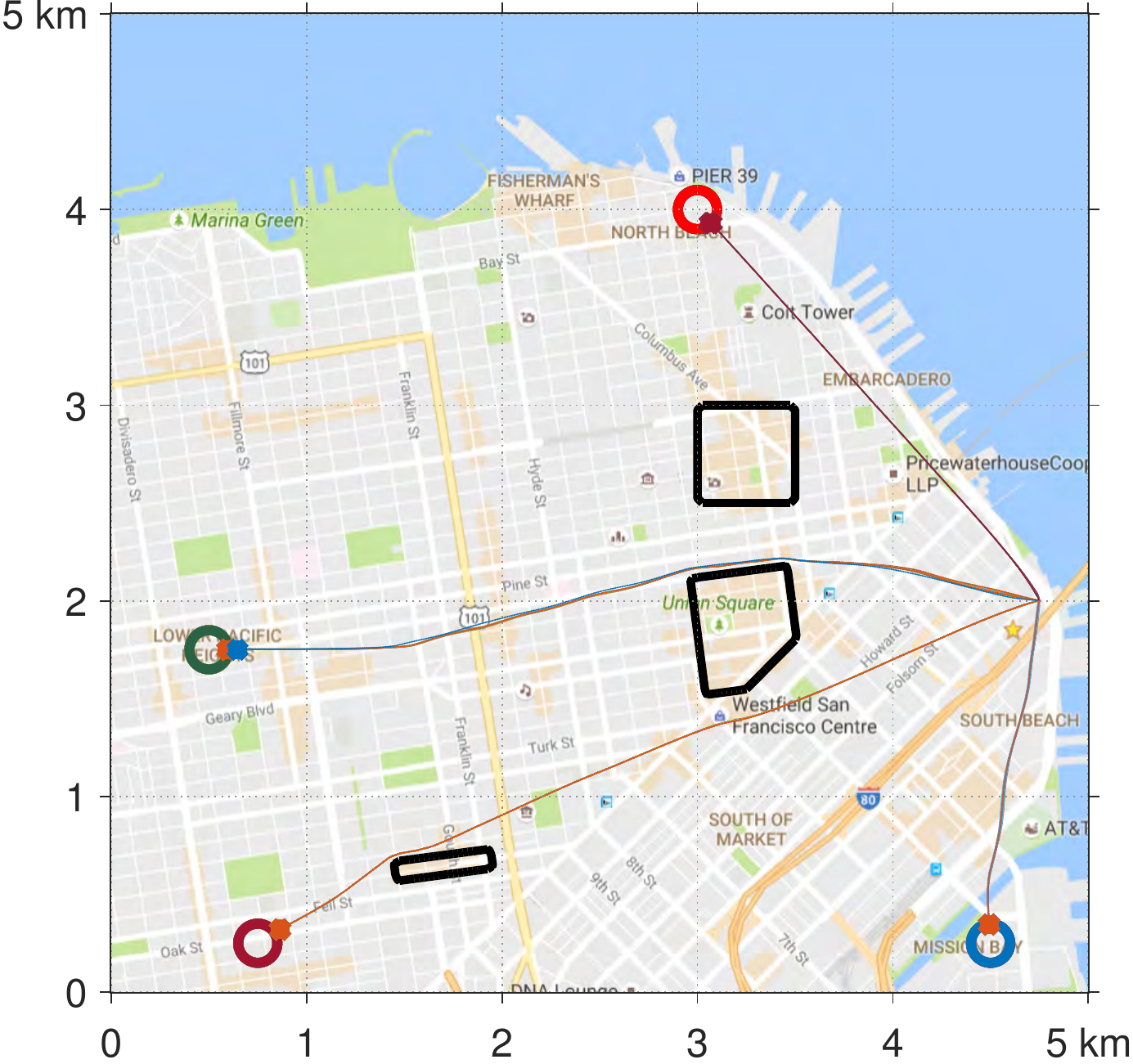}
    \subcaption{6 m/s wind, medium UAV density}
    \label{fig:sf_d6sep5}
  \end{subfigure}~%
  \begin{subfigure}{0.48\columnwidth}
    \includegraphics[width=\columnwidth]{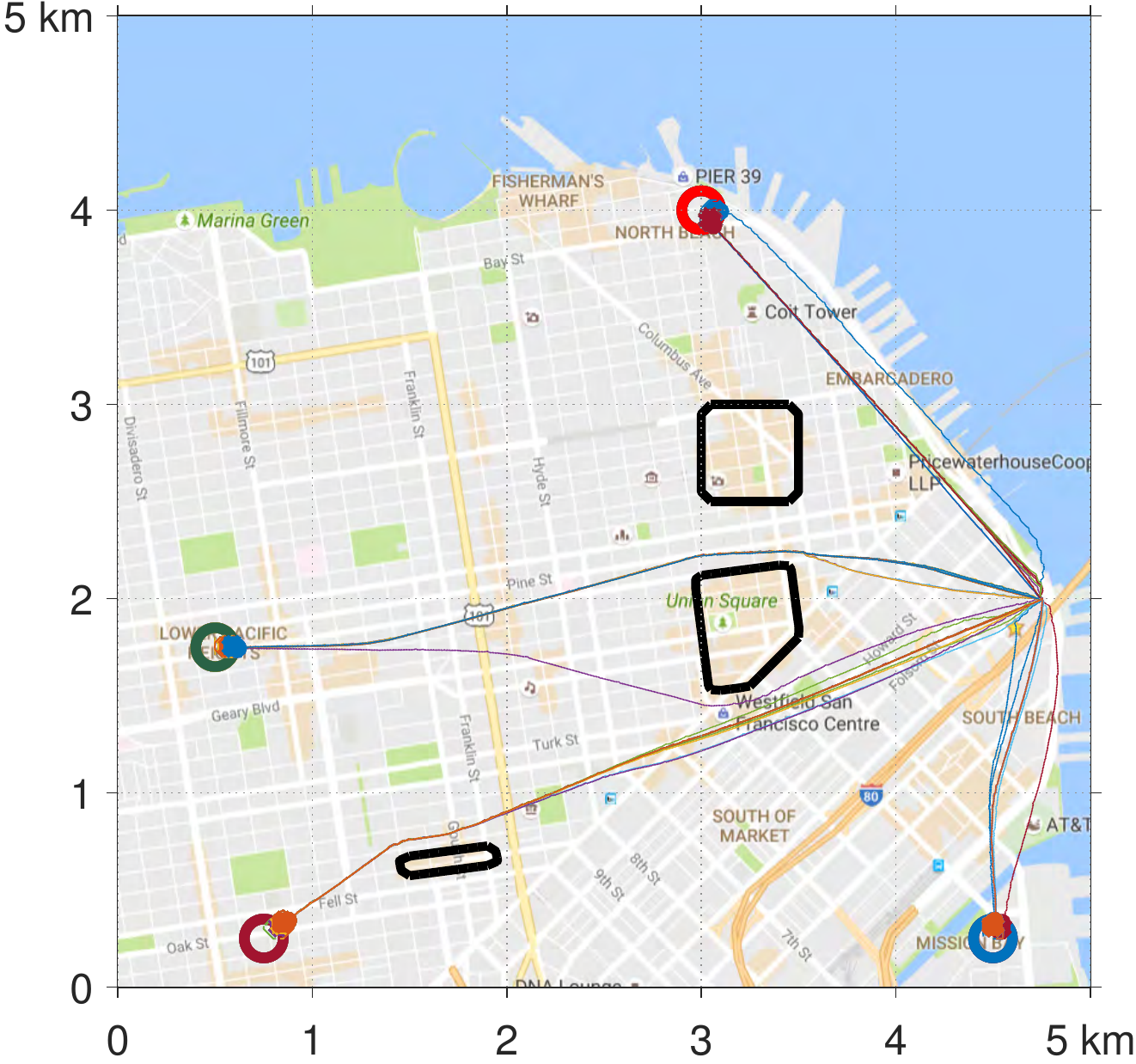}
    \subcaption{11 m/s wind, medium UAV density}
    \label{fig:sf_d11sep5}
  \end{subfigure}%
  \caption{Natural lane forming of UAVs due to disturbance rejection and arrival time constraints. Figures taken from \cite{Chen2017a}.}
  \label{fig:trajectories_sf}
\end{figure}

The trajectory planning of large-scale multi-robot systems has been addressed in work on sequential path planning \cite{Chen2016d}, which robustly synthesizes controllers for many vehicles to reach their destinations while avoiding collisions under the presence of disturbances and a single intruder vehicle. Although reachability is well-suited for these robustness requirements, simultaneous analysis of all vehicles is intractable. Instead, vehicles are assigned a strict priority ordering, with lower-priority vehicles treating higher-priority vehicles as moving obstacles. Robust path planning around these induced obstacles is done using a novel time-varying formulation of reachability \cite{Fisac15}. The result is a reserved ``space-time'' in the airspace for each vehicle, which can be used as a ``last-mile'' solution for getting from air highways to a final postal address. The space-time reservation is dynamically feasible to track even when the vehicle experiences disturbances and performs collision avoidance against an adversarial intruder. Simulations of the robust SPP method over San Francisco for different combinations of wind speeds and UAV densities are shown in Figure \ref{fig:trajectories_sf}. Details can be found in \cite{Chen2016d,Chen2017a,Bansal2017}.

\subsection{Multi-Vehicle Coordination Using HJ Reachability and High-Level Logic}
\begin{figure}[H]
  \centering
  \begin{subfigure}[t]{0.48\columnwidth}
    \includegraphics[width=\columnwidth]{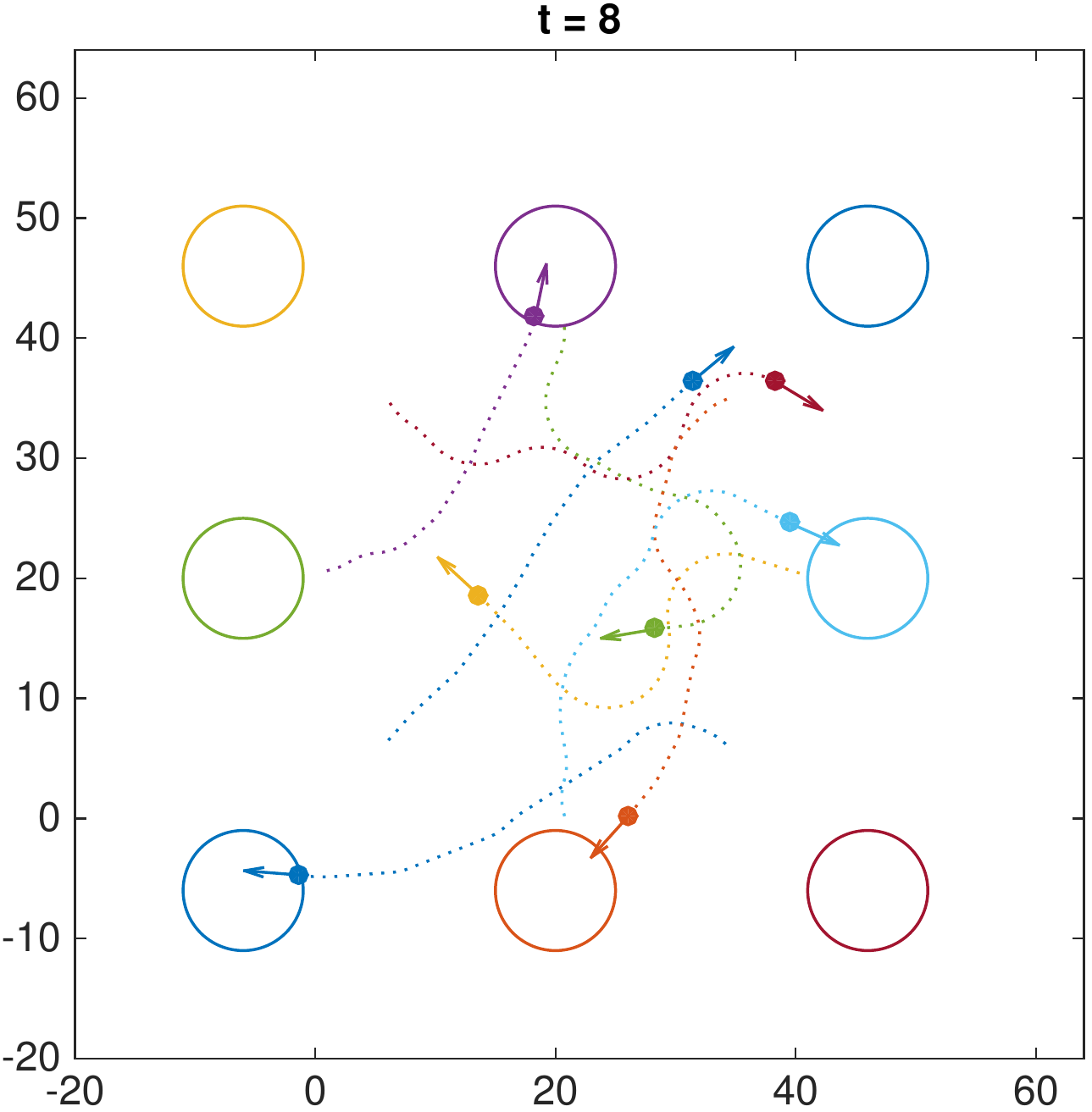}
    \caption{Multi-vehicle collision avoidance simulation. Figure taken from \cite{Chen2016}. \label{subfig:MIP_CA}}
  \end{subfigure}~~%
  \begin{subfigure}[t]{0.48\columnwidth}
    \includegraphics[trim={0 0 15.2cm 0.5cm},clip,width=\columnwidth]{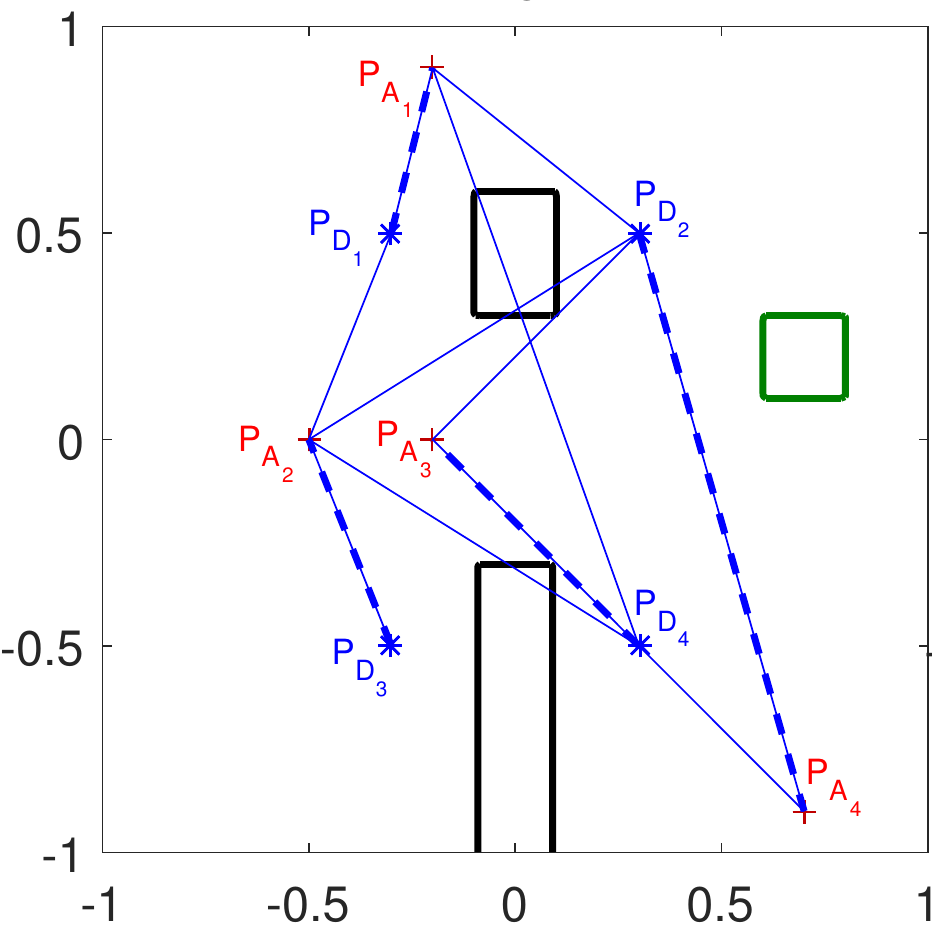}
    \caption{The maximum matching process for rogue UAV interception. Figure taken from \cite{Chen17}.  \label{subfig:MRAGs}}
  \end{subfigure}
  
  \caption{Multi-vehicle analysis using HJ reachability and higher-level logic.}
\end{figure}

In \cite{Chen2016,Chen17}, the scalability limitations of HJ reachability are overcome by a mixed integer program that exploits the properties of pair-wise HJ solutions to provide higher-level control logic. This logic is applied in a couple of different contexts. First, safety guarantees for three-vehicle collision avoidance is proved -- a previously intractable task for HJ reachability -- without incurring significant additional computation cost \cite{Chen2016}. The collision avoidance protocol method is also scalable beyond three vehicles and performs significantly better by several metrics than an extension of pairwise collision avoidance to multi-vehicle collision avoidance. Figure \ref{subfig:MIP_CA} shows an 8-vehicle collision avoidance simulation.

Second, in multiplayer reach-avoid games, two teams of cooperative players with conflicting and asymmetric goals play against each other on some domain, possibly with obstacles. The attacking team tries to arrive at some arbitrary target set in the domain, and the defending team seeks to prevent that by capturing attackers. Such a scenario is useful for intercepting ``rogue" UAVs trying to enter restricted areas of the airspace. The joint solution to this problem is intractable, so a maximum matching approach is taken instead. To each defender, the maximum matching process tries to assign an attacker who is guaranteed to lose to the defender, and the team of defenders coordinate the vehicle-to-vehicle defense. As a result, an upper bound on the number of attackers that can reach the target set can be obtained \cite{Chen17}. The maximum matching result for a particular game setup is shown in Figure \ref{subfig:MRAGs}.
​

\section{Conclusions}
Hamilton-Jacobi (HJ) reachability is a useful tool for guaranteeing goal satisfaction and safety under controlled safety-critical scenarios with bounded disturbances. However, a direct application of HJ reachability in most cases becomes intractable due to its exponentially-scaled computational complexity with respect to the continuous state dimension. In this tutorial, we start from a comprehensive overview of HJ reachability theory from its roots in differential games theory. We then provide an overview of the recent theoretical work that aims at alleviating the curse of dimensionality, including several applications that leverage these ideas to ensure safety.

\section*{ACKNOWLEDGMENT}
The authors would like to thank Jaime F. Fisac whose write up on differential games was immensely helpful in preparing this tutorial document.  

\section{Appendix: Quick-Start Guide}

To familiarize ourselves with the tools available in \textit{toolboxLS} and \textit{helperOC}, we will walk through a simple example file to run several different forms of reachability analysis for a 3D Dubins car example. 

\subsection{Defining and Handling Dynamic Systems\label{sec:dynamic_class}}
Before setting up the analyses, we must first understand how to use the code to create and handle dynamic systems like the 3D Dubins car. In \textit{helperOC} we use object-oriented code to define our dynamics. This allows us to create, for example, a Dubins car ``object" that inherits the properties and functions related to a dynamic system and its own dynamics and parameters. In this section we will review the class structure.

The dynamic systems class is found under \mbox{\textit{helperOC/dynSys/@DynSys}}. This class defines several properties and functions inherent to any dynamical system used for reachability analysis. All systems in \textit{helperOC} are sub-classes of \textit{@DynSys}. The sub-classes are also found in \textit{helperOC/dynSys}. For now we will review the sub-class \textit{@DubinsCar} as an example. This folder contains four files: \textit{DubinsCar.m, dynamics.m, optCtrl.m}, and \textit{optDstb.m.} The dynamics of the Dubins car is defined by:
\begin{equation}
\begin{array}{c}
\left[
\begin{array}{c}
\dot{p_x}\\
\dot{p_y}\\
\dot{\theta}\\
\end{array}
\right]
=
\left[
\begin{array}{c}
v\cos(\theta) + \dstb_1\\
v\sin(\theta) + \dstb_2\\
\ctrl + \dstb_3\\
\end{array}
\right]\\
\\
\ctrl \in \cset , \hspace{.25cm} \dstb=[\dstb_1, \dstb_2, \dstb_3] \in \dset
\end{array}
\end{equation}
\noindent where $\ctrl$ is the control, and $\dstb$ is the disturbance. \mbox{\textit{DubinsCar.m}} is the main function of the Dubins car sub-class. This function defines the properties of a Dubins car (e.g. speed $v$, angular control $\ctrl$, and disturbance $\dstb$), as well as the constructor function for creating a Dubins car object. This function takes in the object parameters and constructs a Dubins car object with said parameters.

The function \textit{dynamics.m} sets the dynamics for the system. Open this file for a demonstration of how to incorporate the dynamics of your system into the code. Note that the inputs are the object, state, control, and disturbance. For time-varying systems, a time input can be included as well.

The functions \textit{optCtrl.m} and \textit{optDstb.m} are used to find the optimal control and disturbance at every grid point in the state space for each time step. These functions are determined by taking the inner product between the spatial gradients of the value function and the system dynamics, i.e. by computing the Hamiltonian, as required in equation (\ref{eqn:hamil_brs}), and as of now must be defined by hand. The control that either maximizes or minimizes the Hamiltonian (depending on what is desired) is the optimal control. The optimal disturbance does the opposite. As an example we will derive the optimal control and disturbance for the Dubins car in the case where uMode = `min' and dMode = `max.' Note that we will use $\valfunc_{p_x}$ to denote the partial derivative of the value function with respect to state $p_x$.
\begin{equation}
\begin{array}{c}
\nabla\valfunc \cdot \fdyn(\state,\ctrl,\dstb) \\
=\valfunc_{p_x}(v\cos(\theta) + \dstb_1)+\valfunc_{p_y}(v\sin(\theta) + \dstb_2)+\valfunc_\theta(\ctrl + \dstb_3)\\
a \in [a_{\min},a_{\max}], [b_1,b_2,b_3] \in [b_{\min},b_{\max}]
\end{array}
\end{equation}

Gathering terms multiplied by the control, we can find the optimal control by taking the argmin of these terms.

\begin{equation}
\begin{array}{c}
\ctrl^* = \arg\min_{\ctrl} \left\{<\nabla\valfunc,\fdyn(\state,\ctrl,\dstb)>\right\} 
=	\arg\min_{\ctrl} \left\{\valfunc_\theta*\ctrl\right\}  \\
=\mbox{$\ctrl_{\min}$ if $\valfunc_\theta\geq 0$, $\ctrl_{\max}$ if $\valfunc_\theta \le 0$}
\end{array}
\end{equation}

We follow a similar procedure to find one of the optimal disturbances.

\begin{equation}
\begin{array}{c}
\dstb_1^* = \arg\max_{\dstb_1} \left\{<\nabla\valfunc,\fdyn(\state,\ctrl,\dstb)>\right\}  \\
=\arg\max_{\dstb_1} \left\{\valfunc_{p_x}*\dstb_1\right\}  \\
=\mbox{$\dstb_{1\max}$ if $\valfunc_{p_x}\geq 0$, $\dstb_{1\min}$ if $\valfunc_{p_x} \le 0$}
\end{array}
\end{equation}

The optimal disturbances $\dstb_2^*$ and $\dstb_3^*$ can be similarly computed. These results are coded into \textit{optCtrl.m} and \textit{optDstb.m} for the Dubins car sub-class. For examples using more complicated dynamics, one can explore other sub-classes within \textit{helperOC/dynSys}.\\
\\

\subsection{Reachability Analysis Setup}
\begin{figure}[]
	\centering
	\includegraphics[width=0.7\columnwidth]{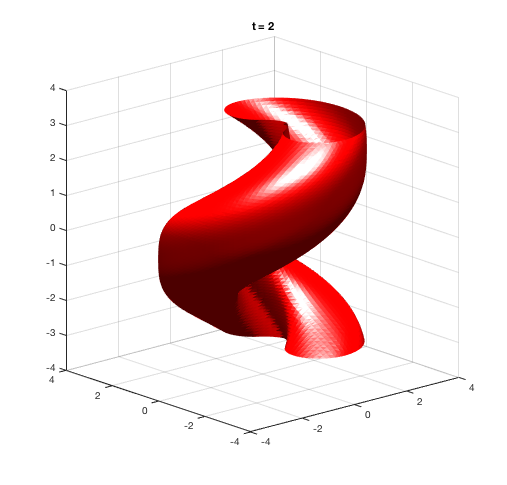}
	\caption{Visualization that should appear when running \textit{tutorial\_test.m}, illustrating a backward reachable set (BRS) for a Dubins car.}
	\label{fig:tutorial_test_image}
\end{figure}

The example file we will use here to define the reachability analysis is \textit{tutorial\_test.m}, and is contained within the helperOC repository. This function consists of modifiable code to run several different forms of reachability analysis for a 3D Dubins car example. Try running this function to verify correct installation. A visualization of a spiral red set in 3D should appear, as shown in Fig. \ref{fig:tutorial_test_image}.

The comments at the top of the script explain how to modify the function to test different versions. In this section we will briefly go through the different blocks of the code in this file.

\begin{enumerate}
	\item \textit{Trajectory Computation?}\\
	This block is set to true when you want to test the results of the reachability analysis using a test trajectory. Note that this can only be used for backward reachable sets and tubes.
	\item \textit{Grid}\\
	In order to compute a reachable set numerically, the level set toolbox discretizes the state space and solves for the value function over a discrete grid.\footnote{This is the key cause of the curse of dimensionality in the BRS computation.} This block defines the grid by setting the minimum and maximum states, along with the number of grid points in each dimension. Periodic dimensions should be noted (to account for periodic behavior), and the grid is created.  Note that the grid bounds should be large enough to enclose the target and the reachable set or tube. Also note that a finer discretization will lead to more accurate results.
	\item \textit{Target Set}\\
	Here we define the target set of the system. As noted in Section \ref{sec:brs}, this is either a subset of the state space we want the system to reach, or an unsafe set that we want the system to avoid. In this example we use the function \textit{shapeCylinder.m} to create a target set that is circular in $p_x,p_y$ space and encompasses all $\theta$ states. Functions for other shapes can be found in the \textit{toolboxLS} user manual.
	\item \textit{Time Vector}\\
	In this block the initial and final times are set, as well as the time step desired. Note that for forward reachable sets the variable \textit{tau} moves forward in time, and for backward reachable sets \textit{tau} moves backward in time.  See Section \ref{sec:frs} for more details on FRSs and BRSs.
	\item \textit{Problem Parameters}\\
	Here the problem parameters for the dynamical system are introduced. These problem parameters are defined by the class of the dynamical system (in this case, a Dubins car). The control and (if applicable) disturbance modes (\textit{uMode} and \textit{dMode}) are also defined here. This refers to whether the control (or disturbance) is trying to maximize or minimize the value function (see Section \ref{sec:minmax} for details). Table \ref{Table:uMode} shows the modes needed depending on the reachability problem. The disturbance mode \textit{dMode} is generally the opposite of \textit{uMode} for the worst-case analysis.  The table also differentiates between whether the control is trying to reach the target set (goal), or avoid the target set (avoid).
	\begin{table}[h]
		\centering
		\caption{\textit{uMode} Conditions}
		\label{Table:uMode}
		\begin{tabular}{l|l|l|}
			\cline{2-3}
			& \multicolumn{2}{c|}{\textit{uMode}}                                                   \\ \hline
			\multicolumn{1}{|l|}{Target set} & goal (larger set) & avoid (smaller set) \\ \hline
			\multicolumn{1}{|l|}{Forward}    & max                                  & min                                   \\ \hline
			\multicolumn{1}{|l|}{Backward}   & min                                  & max                                   \\ \hline
		\end{tabular}
	\end{table}
	\item \textit{Pack Problem Parameters}\\
	This block packs problem parameters into the variables needed for the reachability computation. The dynamical system is defined using the input parameters from the previous block and calling upon the appropriate dynamic system class that was created in Section \ref{sec:dynamic_class}. The system, grid, \textit{uMode} (and \textit{dMode} if applicable), and accuracy level are set. The accuracy options are \textit{low}, \textit{medium}, \textit{high}, and \textit{veryHigh}. Note that higher accuracy results in a more accurate gradient calculation of the value function, but takes more time to compute the value function.
	\item\textit{ Obstacles}\\
	Obstacles (or unsafe sets) should be defined here using the same format used for creating the target set. The obstacles should then be combined in a cell structure and set to \textit{HJIextraArgs.obstacles}.
	\item \textit{Compute Value Function}\\
	In this block we set the \textit{HJIextraArgs} parameter to visualize the reachability analysis during computation. We then use the main function of \textit{helperOC}, \textit{HJIPDE\_solve.m}, to perform the reachability analysis and to acquire the discrete form of the continuous value function in equation (\ref{eqn:val_fn}). Note that the function can solve for a reachable set by setting the \textit{minWith} input to `none', or a tube by setting the \textit{minWith} input to `zero'.  The differences between sets and tubes are explained in Section \ref{sec:brs_brt}. More information on \textit{HJIPDE\_solve.m} and extra functionalities are in Section \ref{sec:hjipde_solve}.
	\item \textit{Compute Optimal Trajectory for Some Initial State}\\
	If the Trajectory Computation block is set to true, this block computes and visualizes an optimal trajectory from a given initial state and the optimal controller derived from the value function, which is computed using equation (\ref{eq:OptCtrl_brs}) for a BRS, for example.\\
\end{enumerate}

\subsection{Using \textit{HJIPDE\_Solve.m}\label{sec:hjipde_solve}}
The main function used by \textit{helperOC} is \textit{HJIPDE\_Solve.m}, which can be found in \textit{helperOC/valFuncs}. This function interfaces the tools developed in helperOC with the functions used in toolboxLS. The inputs are the initial values at each grid point (\textit{data0}), the time vector (\textit{tau}), the problem parameters for toolboxLS (\textit{schemeData}), whether to compute a set or a tube (\textit{minWith}), and any additional inputs desired (\textit{extraArgs}). The outputs are the value function at each grid point at each time step (\textit{data}), the time vector (\textit{tau}), and any additional outputs desired (\textit{extraOuts}).

The range of possibilities for the \textit{extraArgs} input are described in comments at the beginning of the function. You can include obstacles, visualize the set over time, stop when the set reaches some initial state, save the data periodically, and more.\\
\\


\bibliographystyle{IEEEtran}
\bibliography{references}
\end{document}